\documentclass[aps,pre,twocolumn,showpacs,groupeaddress]{revtex4}
\usepackage{braket}
\usepackage{comment} 
\usepackage{graphicx}
\usepackage{dcolumn}
\usepackage{bm}

\begin{document}
\title{Quantum fidelity approach to the ground state properties of 
the 1D ANNNI model in a transverse field}
\author{O. F. \surname{de Alcantara Bonfim}}
\email{bonfim@up.edu}
\affiliation{Department of Physics, University of Portland, Portland, Oregon 97203, USA}
\author{B. \surname{Boechat}}
\email{bmbp@if.uff.br}
\affiliation{Departamento de F\'{\i}sica, Universidade Federal Fluminense\\
Av. Litor\^anea s/n,  Niter\'oi, 24210-340, RJ, Brazil}
\author{J. \surname{Florencio}}
\email{jfj@if.uff.br}
\affiliation{Departamento de F\'{\i}sica, Universidade Federal Fluminense\\
Av. Litor\^anea s/n,  Niter\'oi, 24210-340, RJ, Brazil}


\begin{abstract}
In this work we analyze the ground-state properties of the $s=1/2$ one-dimensional ANNNI 
model in a transverse field  using the quantum fidelity approach. 
We numerically determined the fidelity susceptibility as a function of the transverse 
 field $B_x$ and the strength of the next-nearest-neighbor interaction $J_2$, for
 systems of up to 24 spins.  We also examine the ground-state vector
with respect to the spatial ordering of the spins.
The ground-state phase diagram shows ferromagnetic, 
floating,  $\Braket{2,2}$ phases, and we predict an infinite number of
modulated phases in the thermodynamic limit ($L \rightarrow \infty$).  
Paramagnetism only occurs for larger magnetic fields.
The transition lines separating the modulated phases seem to be of second-order, 
whereas the line between the floating  and the $\Braket{2,2}$ phases is possibly of first-order.

\end{abstract}

\pacs{75.10.Pq,75.10.Jm}

\maketitle

\section {\label{sec:I} Introduction}

At very low temperatures, quantum fluctuations play an important role 
in the characterization of the ground-state properties of quantum 
systems~\cite{Sac99}. 
These fluctuations are induced by varying the relative strength 
of competing interactions among the constituents of the system 
or by changing the strength of the applied fields.  
When large enough,  quantum fluctuations dramatically 
change the nature of a given ground-state. 
A quantum phase transition may occur, thereby creating a boundary 
between distinct ground-states.

The one-dimensional axial next-nearest neighbor Ising (1D ANNNI) model in
a transverse field is one of the simplest models in which competing interactions 
lead to modulated magnetic orders, frustration, commensurate-incommensurate 
transitions, etc. These features are known to appear in the ground-state of the 
model in the one-dimensional case.

Frustration in the 1D ANNNI model arises from the competition between 
nearest-neighbor interactions which favor ferromagnetic alignment
of neighboring spins, while an interaction with opposite sign between 
the next-nearest-neighbors fosters antiferromagnetism.
At $T = 0$, the presence of a transverse magnetic field gives rise 
to quantum fluctuations that play an analogous role as that of temperature 
in thermal magnetic systems that are responsible for triggering 
phase transitions. 
 
 In one dimension, the ANNNI model in a transverse field is actually
an extension of the transverse Ising model.
The latter consists of Ising spins with nearest-neighbor 
interactions in the presence of a magnetic field in the transverse direction. 
The transverse Ising model was initially used to explain the order-disorder 
transitions observed in KDP ferroelectrics~\cite{Gen63}. 
An experimental realization of that model in real magnetic 
systems was observed in LiHoF${_4}$ in an external field~\cite{Bit96}. 
An exact solution to the model in one dimension was subsequently 
found by Pfeuty by mapping the set of the original spin operators onto a 
new set of noninteracting spinless Fermi operators~\cite{Pfe70}.  
 Recently, a degenerate Bose gas of rubidium confined in a tilted optical 
 lattice was used to simulate a chain of interacting Ising spins 
 in the presence of both transverse and longitudinal fields~\cite{Sim11}.  
 It has also been proven that the ground-state properties of 
 the $d$-dimensional Ising model with a transverse field, is equivalent 
 to the $(d + 1)$-dimensional Ising model without a magnetic field 
 at finite temperatures~\cite{You75,Her76,Suz76}. 
 
In the case of the 1D ANNNI model in a transverse field at $T=0$ and 
the 2D ANNNI model (without transverse field) at finite T, such equivalence 
may only exist in the limit of very strong transverse field and in the weak-coupling 
limit of the NN- and NNN-interactions of the 1D model \cite{Ruj81,Bar81,Bar82}.
There is no guarantee that the ground-state phase diagrams 
of those models bear any resemblances to each other. Therefore we shall not compare
the phase diagrams of these two models in this work.

The transverse 1D ANNNI model has been the subject of great
interest ~\cite{Suz13,Dut15}, in part due to the number of
quantum phases with unusual and intriguing features
it displays.  Several analytical and numerical
methods have been employed to establish its phase
diagram.  Among those studies, there are analysis
using quantum Monte Carlo~\cite{Ari91},
exact diagonalization of small lattice systems~\cite{Sen92,Rie96}, 
interface approach~\cite{Sen97},  scaling behavior
of the energy gap~\cite{Gui02}, bosonization and renormalization groups 
methods~\cite{Dut03},  density matrix renormalization 
group~\cite{Bec06,Bec07}, perturbation theory~\cite{Cha07}, 
and matrix product states~\cite{Nag11}.

The phase diagrams from those works do not necessarily
agree with each other.  In the following we
discuss the common features as well as some of the
differences between them.
In most of the studies, there is ferromagnetism for
$J_2 < 0.5$ and $\Braket{2,2}$ antiphase for $J_2 > 0.5$.
The transition lines usually end at the multicritical point $(J_2,B_x)=(0.5,0.0)$.
The phase diagram of Dutta and Sen shows antiferromagnetism instead of
the  $\Braket{2,2}$ antiphase for $J_2 > 0.5$~\cite{Dut03}.  That is a rather
surprising result not to show the antiphase, since even in the classical case,
$B_x = 0$, that antiphase is energetically favorable.
Some authors obtain diagrams with 5 phases, namely, ferromagnetic, paramagnetic, modulated paramagnetic, 
floating, and antiphase.  Such are the diagrams of Arizmendi et al.~\cite{Ari91}, Sen et al.~\cite{Sen92}, 
and Beccaria et al.~\cite{Bec06,Bec07}.   
On the other hand, Rieger and Uimin~\cite{Rie96}, Chandra and Dasgupta~\cite{Cha07}, and  Nagy~\cite{Nag11} 
present diagrams with 4 phases, ferromagnetic, paramagnetic, floating, and antiphase.
In Refs.\cite{Rie96} and \cite{Nag11} the boundary lines meet at the multicritical point, 
whereas in Ref.~\cite{Cha07} the paramagnetic phase
is restricted to suficiently high $B_x$, thus its boundary lines do not reach the multicritical point.
In the studies by Sen~\cite{Sen97} and Guimar\~aes et al.~\cite{Gui02}, one finds 
diagrams with 3 phases only, ferromagnetic, paramagnetic, and antiphase,
where their transition lines end at the multicritical point.
The phase diagram of Dutta and Sen~\cite{Dut03} displays
ferromagnetism, a spin-flop phase, a floating phase, and
an antiferromagnetic phase.  In that work, the floating phase
lies between the antiferromagnetic and the spin-flop phases.
Such spin-flop and antiferromagnetic  phases
do not appear in any of the other phase diagrams  in the literature.  
In addition, their transition lines do not end at the multicritical point.
As one can see, there is not a consensus on the ground-state
phase diagram of the model.  The number,  nature, or location 
of the phases usually vary from one work to another.
In any case, all the studies in the literature report
on a finite number of phases.  As we shall see below, our
phase diagram agrees with some of the works  in the literature 
with regard to the existence of ferromagnetic, floating, and the 
antiphase.  However, our numerical results suggest that there are
an infinite number of modulated phases between the ferromagnetic
and the floating phase.   Such scenario is similar
 to the one found in the the work of Fisher and Selke~\cite{Fis80}
 on the low-temperature phase diagram of an Ising model 
 with competing interactions. In that study the phase
 diagram shows an infinite number of commensurate phases.

While the identification of the usual thermal phase 
transitions relies mostly on the behavior of an order 
parameter or on an appropriate correlation 
function, quantum phase transitions can also be  
characterized solely by the properties of the ground-state 
eigenvectors of the system on each side of the boundary 
between two competing quantum mechanical states.
We use  fidelity susceptibility to determine the phase boundary lines,
as well as a direct inspection of the eigenvectors to understand the nature of the phases.
In our work, paragnetism only occurs at high fields $B_x$, hence it does
not appear in our phase diagram, which covers the low field region only.
In addition, our numerical analysis  points to the the existence 
of a region of finite width for the floating phase.

\section{\label{sec:II} The Model} 

The one-dimensional ANNNI model in the presence of a 
transverse magnetic field is defined as
\begin{equation}
{\cal H} = -J_1\sum_{i} \sigma^z_{i}\sigma^z_{i+1}
+ J_{2} \sum_{i} \sigma^z_{i}\sigma^z_{i+2} - B_x \sum_{i} \sigma^x_{i}.
\label{Hamilt}
\end{equation}
The system consists of $L$ spins, with $s=1/2$, 
where $\sigma^{\alpha}_i$ $(\alpha = x,y,z)$ is the $\alpha$-component  
of a Pauli operator located at site $i$ in a chain
where periodic boundary conditions are imposed.
We considered ferromagnetic nearest-neighbor Ising coupling $J_1 > 0$
and antiferromagnetic next-nearest-neighbor interaction $J_{2} > 0$.   
$B_x$ is  the strength of a transverse applied magnetic 
 field along the $x$-direction.  
We set $J_1 = 1$ as the unit of energy.

At $T = 0$ and in the absence of an external 
magnetic field ($B_x=0$), the model is trivially solvable and
 presents several ordered phases.
 For $J_2 < 0.5$, the ground-state ordering is ferromagnetic, 
 and for $J_2 > 0.5$, the ordering changes 
 to a periodic configuration with two up-spins followed 
 by two down-spins which is termed the $\Braket{2,2}$-phase,
or antiphase. 
In this work we have used the notation $\Braket{p,q}$ to represent
a periodic phase, with $p$ up-spins followed by $q$ down-spins. 
At  $J_2 = 0.5$, the model has a multiphase point where the 
ground-state is infinitely degenerate  and a large number 
of $\Braket{p,q}$-phases are present, as well as other 
spin configurations. The number of phases increases 
exponentially with the size of the system~\cite{Red81,Sel88}. 
On the other hand, for a non-zero external magnetic field and 
$J_2 = 0$, the model reduces to the Ising model 
in a transverse field, which was solved exactly by Pfeuty~\cite{Pfe70}. 
The transverse magnetic field  
induces quantum fluctuations that eventually drive 
the system through a quantum phase transition. 
Its ground-state undergoes a  second-order quantum phase 
transition at $B_x = 1$, separating ferromagnetic from 
paramagnetic phases.
In the 1D transverse ANNNI model, next-nearest-neighbor 
interactions introduces frustration to the magnetic order. 
A much richer variety of phases becomes 
possible when one varies the strength of the interactions among the spins  
or their couplings to the magnetic field. 

Given that insofar there is not a definite answer to
the problem of the ground-state properties of the
transverse ANNNI model, where different approaches
yield distinct phase diagrams, we use quantum fidelity
method together with direct inspection of the
ground-state eigenvector to shed some light into the
problem. We believe our approach is suitable
 because both the fidelity susceptibility 
and ground-state eigenvector
provide detailed direct information about  boundary and
nature of the ground-state phases.  We investigate
how the phase diagram evolves as we consider larger
and larger lattices.  Our results are consistent with
some known results, such as the classical multicritical
point, the Pfeuty quantum transition 
point, and the exact Peschel-Emery line which runs between
those two points in the phase diagram~\cite{Pes81}.
From our results for finite sized systems we can infer 
which phases will be present in the thermodynamic limit.

\begin{figure}
\includegraphics[width=8.0cm, height= 6.0cm, angle=0]{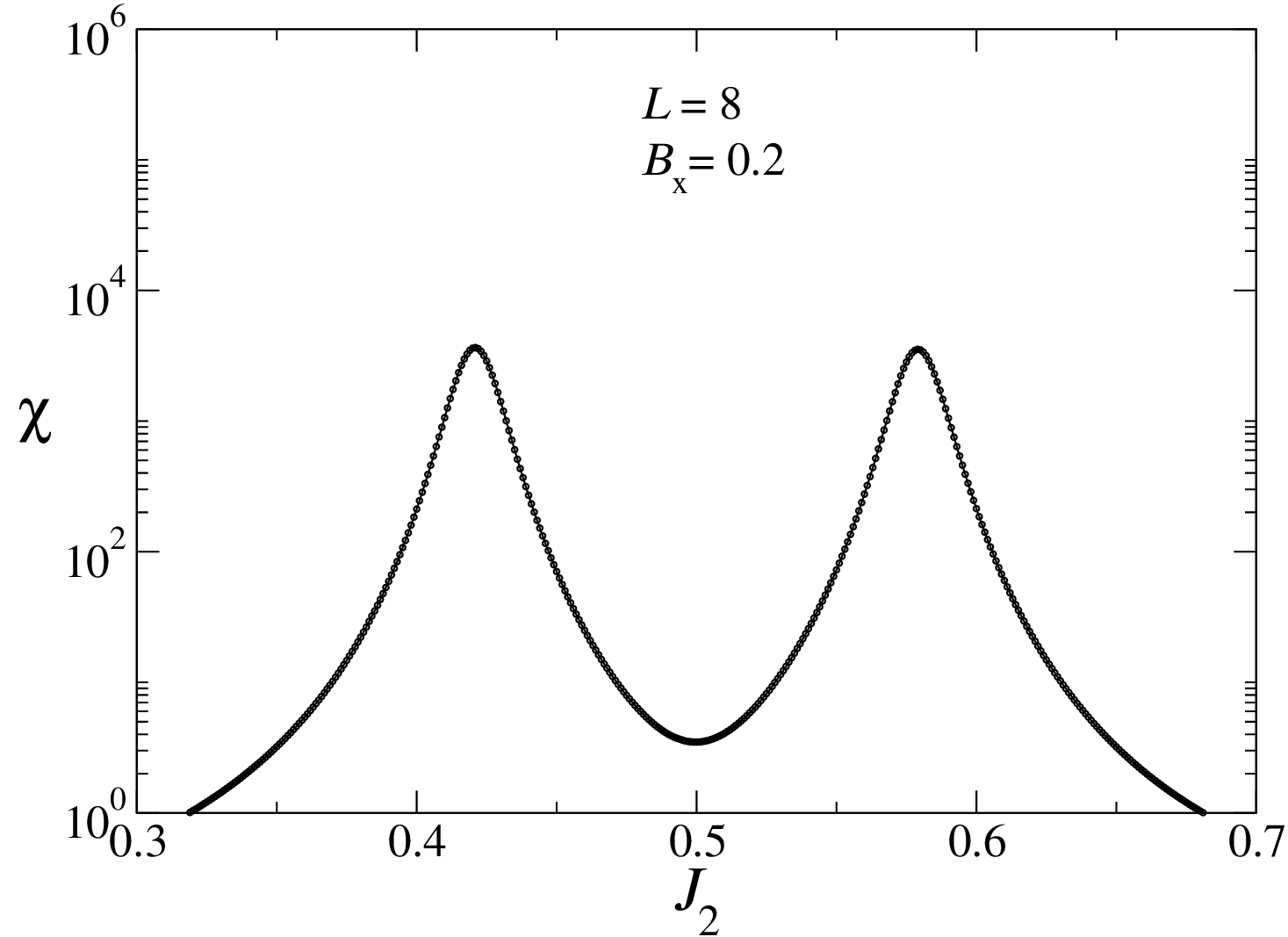}
\setlength{\abovecaptionskip}{0pt}
\setlength{\belowcaptionskip}{20pt}
\caption{Fidelity susceptibility as a function of the 
next-nearest-neighbor coupling $J_2$ for the transverse ANNNI model
with $B_x=0.2$, for the case of a chain with  $L = 8$ spins. 
Here, and also in the next figures, $J_1=1$ is set as the energy unit.
The  locations of the peaks give the transition points. 
\label{fig:fig1}
}
\end{figure}

\begin{figure}
\includegraphics[width=8.0cm, height= 6.0cm, angle=0]{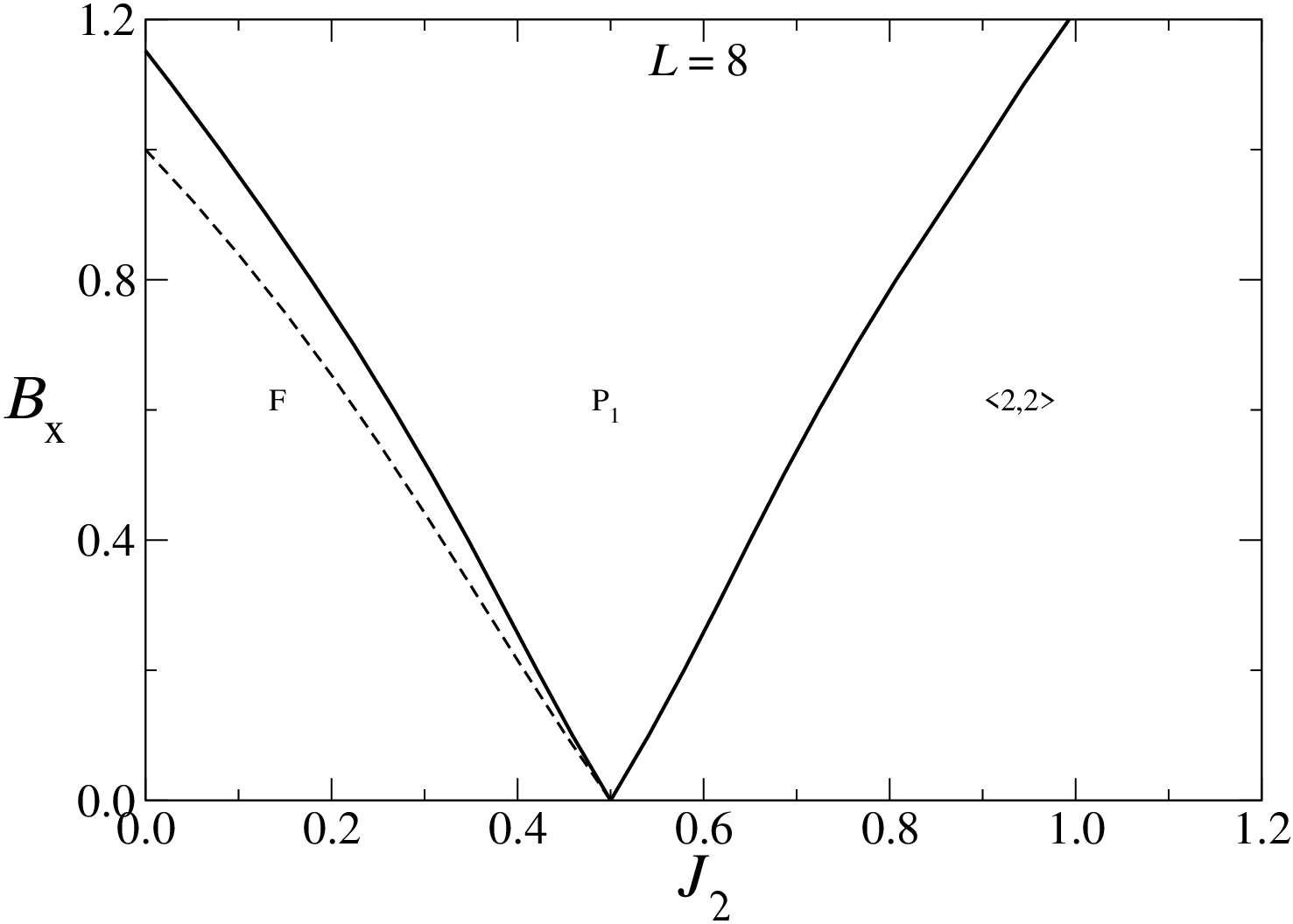}
\setlength{\abovecaptionskip}{0pt}
\setlength{\belowcaptionskip}{00pt}
\caption{Phase diagram in the $(J_2, B_x)$-plane for a system of size $L= 8$ .
The system displays three phase regions, ferromagnetic $F$,
P$_1$, and  $\Braket{2,2}$. No additional phases are present here.
The dashed boundary is the exact Peschel-Emery line.
\label{fig:fig2}
}
\end{figure}

\begin{figure}
\includegraphics[width=8.0cm, height= 6.0cm, angle=0]{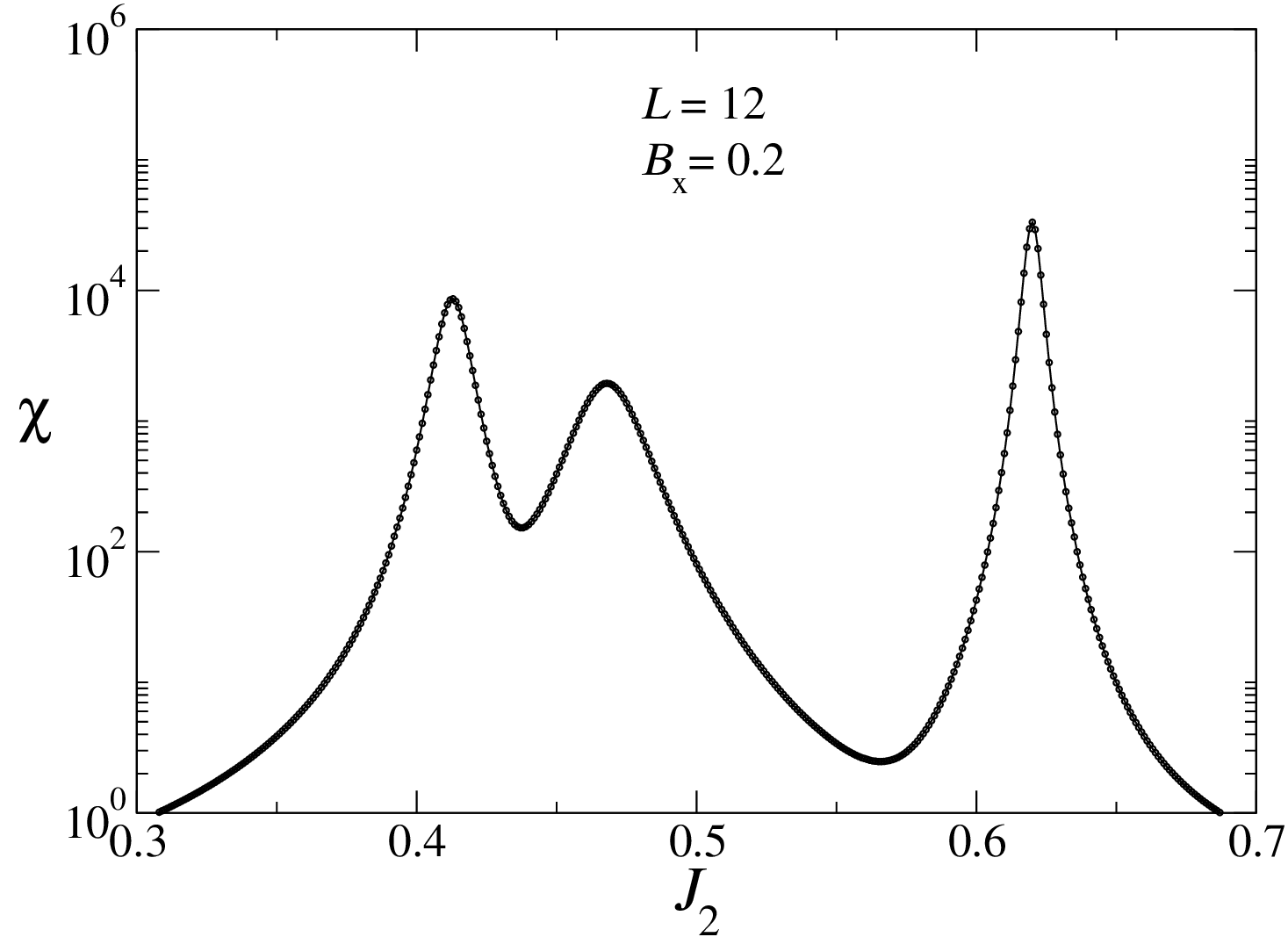}
\setlength{\abovecaptionskip}{0pt}
\setlength{\belowcaptionskip}{30pt}
\caption{Fidelity susceptibility as a function of 
next-nearest-neighbor coupling $J_2$, 
with $B_x=0.2$, for the case  $L = 12$ spins. 
The  locations of the peaks give the transition points. 
\label{fig:fig3}
}
\end{figure}

\begin{figure}
\includegraphics[width=8.0cm, height= 6.0cm, angle=0]{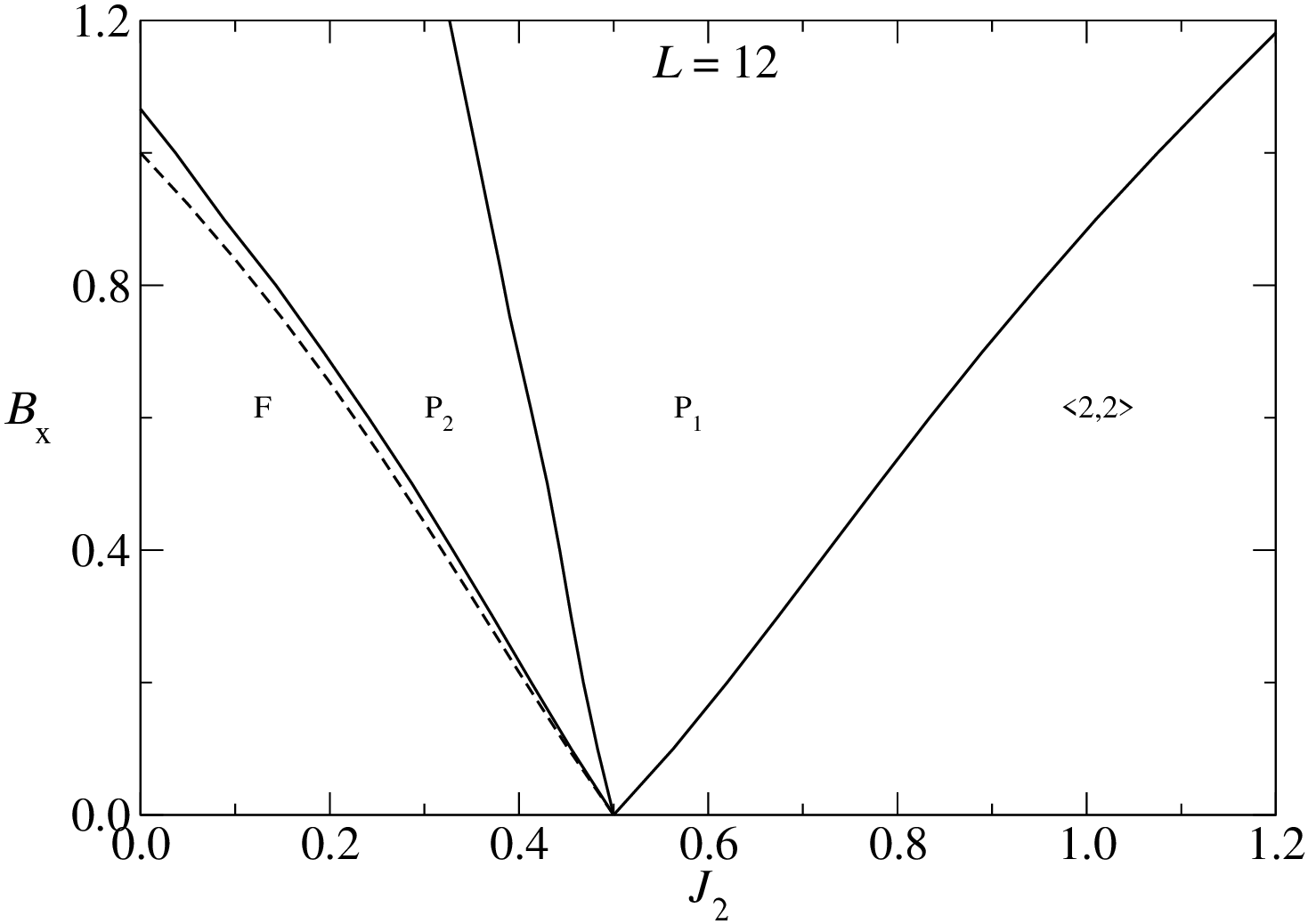}
\setlength{\abovecaptionskip}{0pt}
\setlength{\belowcaptionskip}{0pt}
\caption{Phase diagram in the $(J_2, B_x)$-plane.
The system displays four phase regions, ferromagnetic $F$,
 P$_1$ and $P_2$, and  $\Braket{2,2}$. 
Again, the dashed boundary is the Peschel-Emery line.
\label{fig:fig4}
}
\end{figure}

\begin{figure}
\includegraphics[width=8.0cm, height= 6.0cm, angle=0]{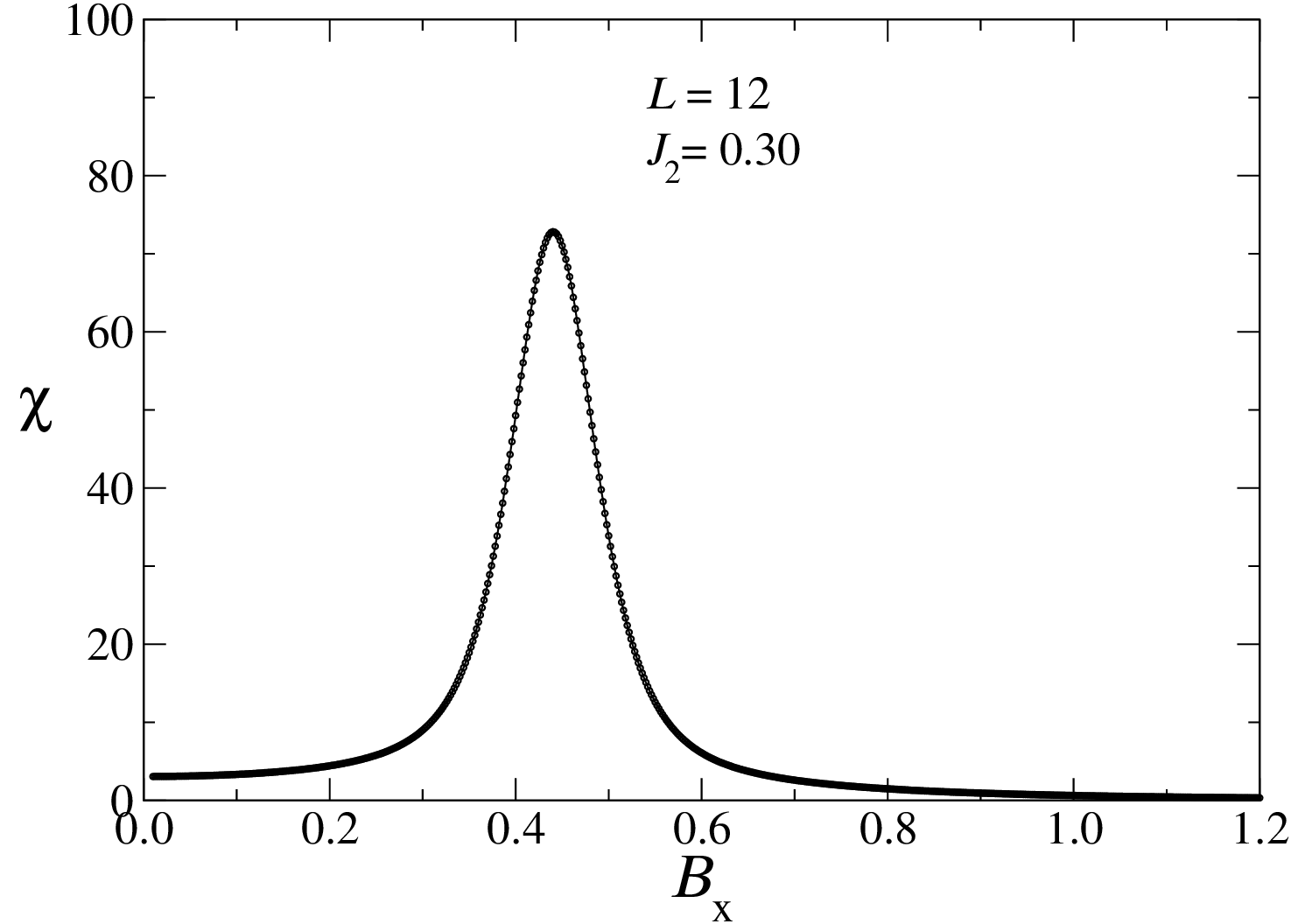}
\setlength{\abovecaptionskip}{0pt}
\setlength{\belowcaptionskip}{40pt}
\caption{Fidelity susceptibility as a function of the transverse field $B_x$
for the transverse ANNNI model with $J_2=0.30$, for the case of a chain 
with  $L = 12$ spins. The  location of the peak gives the transition point. 
\label{fig:fig5}
}
\end{figure}

\begin{figure}
\includegraphics[width=8.0cm, height= 6.0cm, angle=0]{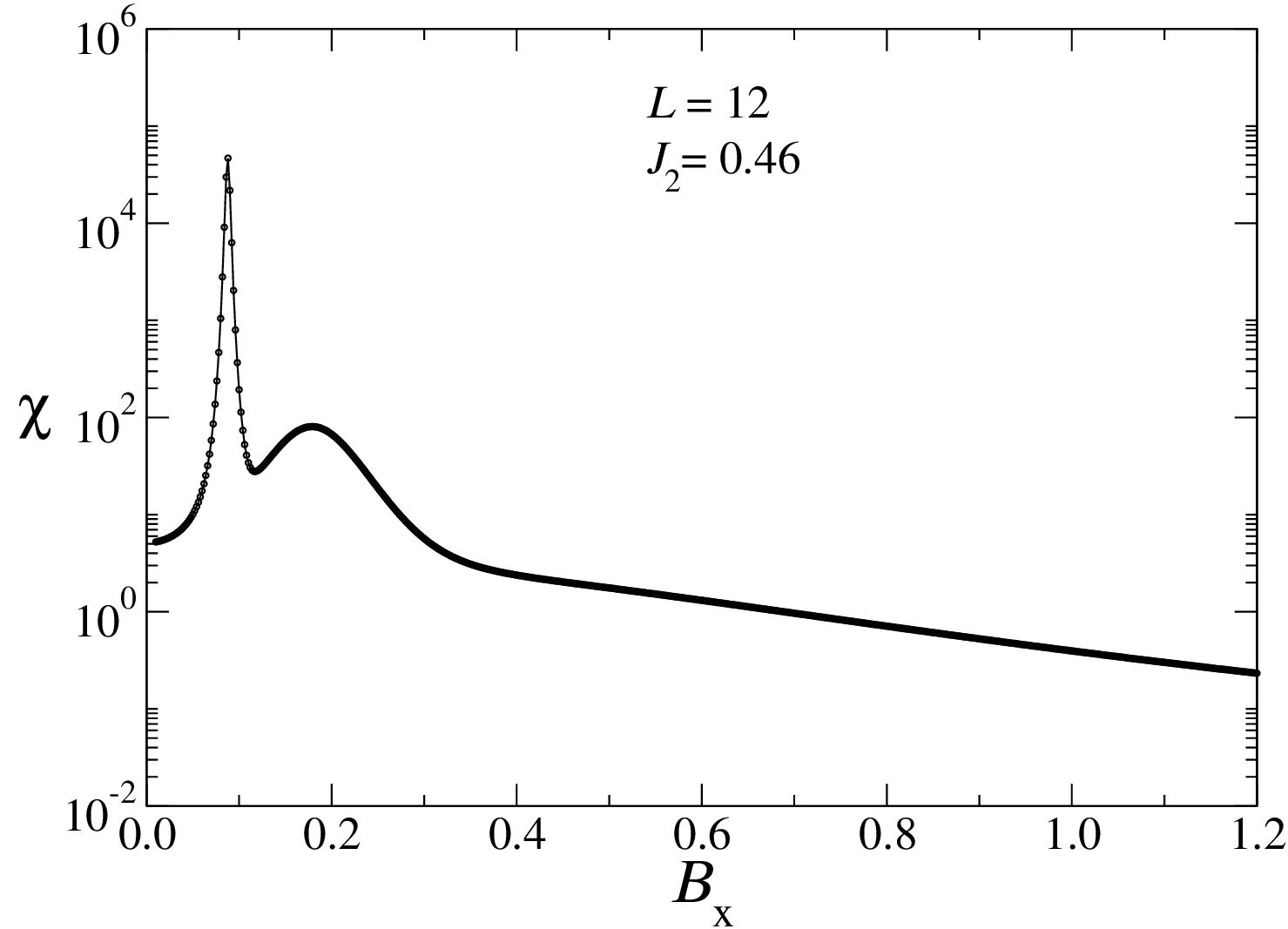}
\setlength{\abovecaptionskip}{0pt}
\setlength{\belowcaptionskip}{30pt}
\caption{Fidelity susceptibility versus the transverse field $B_x$,
for $J_2=0.46$, in the case $L = 12$. The locations of the peaks give the 
transition points. 
\label{fig:fig6}
}
\end{figure}

\begin{figure}
\includegraphics[width=8.0cm, height= 6.0cm, angle=0]{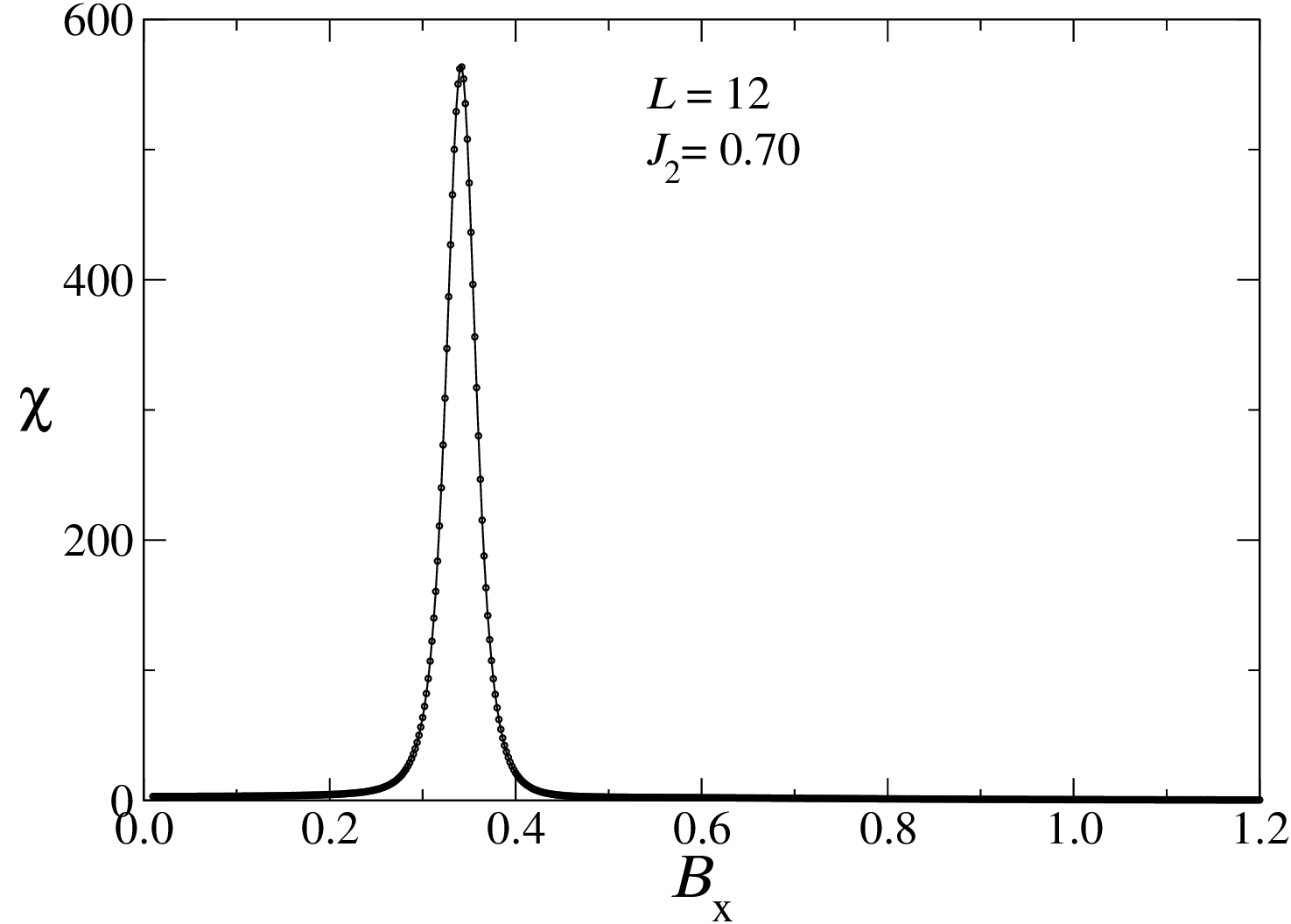}
\setlength{\abovecaptionskip}{0pt}
\setlength{\belowcaptionskip}{70pt}
\caption{Fidelity susceptibility as a function of the transverse field $B_x$,
with $J_2=0.70$, for the case $L = 12$. The  location of the peak gives the transition point. 
\label{fig:fig7}
}
\end{figure}

\section{\label{sec:III} The Fidelity Method}

Suppose the Hamiltonian of the system depends on 
a parameter $\lambda$, which drives the system through 
a quantum phase transition at a critical value 
$\lambda= \lambda_c$. 
Quantum fidelity is defined as the absolute value 
of the overlap between neighboring 
ground-sates of the system~\cite{And67,Zan06},
\begin{equation}
           F(\lambda,\delta) = |\Braket{\psi(\lambda - \delta)|\psi(\lambda + \delta)}|.
\label{fidelity}
\end{equation}
Here $\Ket{\psi}$ is the quantum non-degenerate 
ground-state eigenvector that is evaluated at some 
value of $\lambda$, shifted by an arbitrary small 
quantity $\delta$ around it. 
In addition to the dependence on $\lambda$ and $\delta$, 
the quantum fidelity is also a function of the size of the system. 
The basic idea behind the fidelity approach is that the 
overlap of the ground-state for values of the 
parameter $\lambda$ between the two sides 
of a quantum transition, exhibits a considerable 
drop due to the distinct nature of the ground states 
on each side of the phase boundary. 
Quantum fidelity has been used in quantum information 
theory~\cite{Ben92} as well as in condensed matter 
physics, in particular in the study of topological 
phases~\cite{Aba08,Coz07}.

For a fixed value $L$ and in the limit of very small $\delta$, 
the quantum fidelity may be written as  a Taylor expansion, 
\begin{equation}
             F(\lambda,\delta) = 1 - \chi(\lambda)\delta{^2} +  {\cal{O}}(\delta^4),
\label{susceptibility}
\end{equation}
where the ground-state eigenvector is normalized to unity. 
The quantity $\chi(\lambda)$ is called the fidelity susceptibility 
and will reach a maximum at the boundary between adjacent 
quantum phases. 
We used the fidelity susceptibility to find the 
phase boundary lines the $(J_2,B_x)$-plane and 
compare them with the results obtained by other 
methods.

 To determine the ground-state energy and eigenvector 
as a function of $\lambda$, 
we employed both Lanczos and the conjugate-gradient methods. The latter is
known to be a fast and reliable computational algorithm. 
It has been used in statistical physics, especially 
in the context of Hamiltonian models and of transfer-matrix
 techniques ~\cite{Nig90,Nig93}. Both methods give the same 
ground-state eigenvalues and eigenstates within a given precision. 
Depending on the size of the system, 
the ground-state energy is calculated 
with precision between $10^{-10}$ and $10^{-12}$. 
We have used $\delta = 0.001$ in all calculations 
involving the  fidelity susceptibility.
For the location of each point at the critical 
boundary, we calculated the maximum value 
of the fidelity susceptibility as defined by 
 Eq.~\ref{susceptibility}.

\begin{figure}
\includegraphics[width=8.0cm,height= 6.0cm,angle=0]{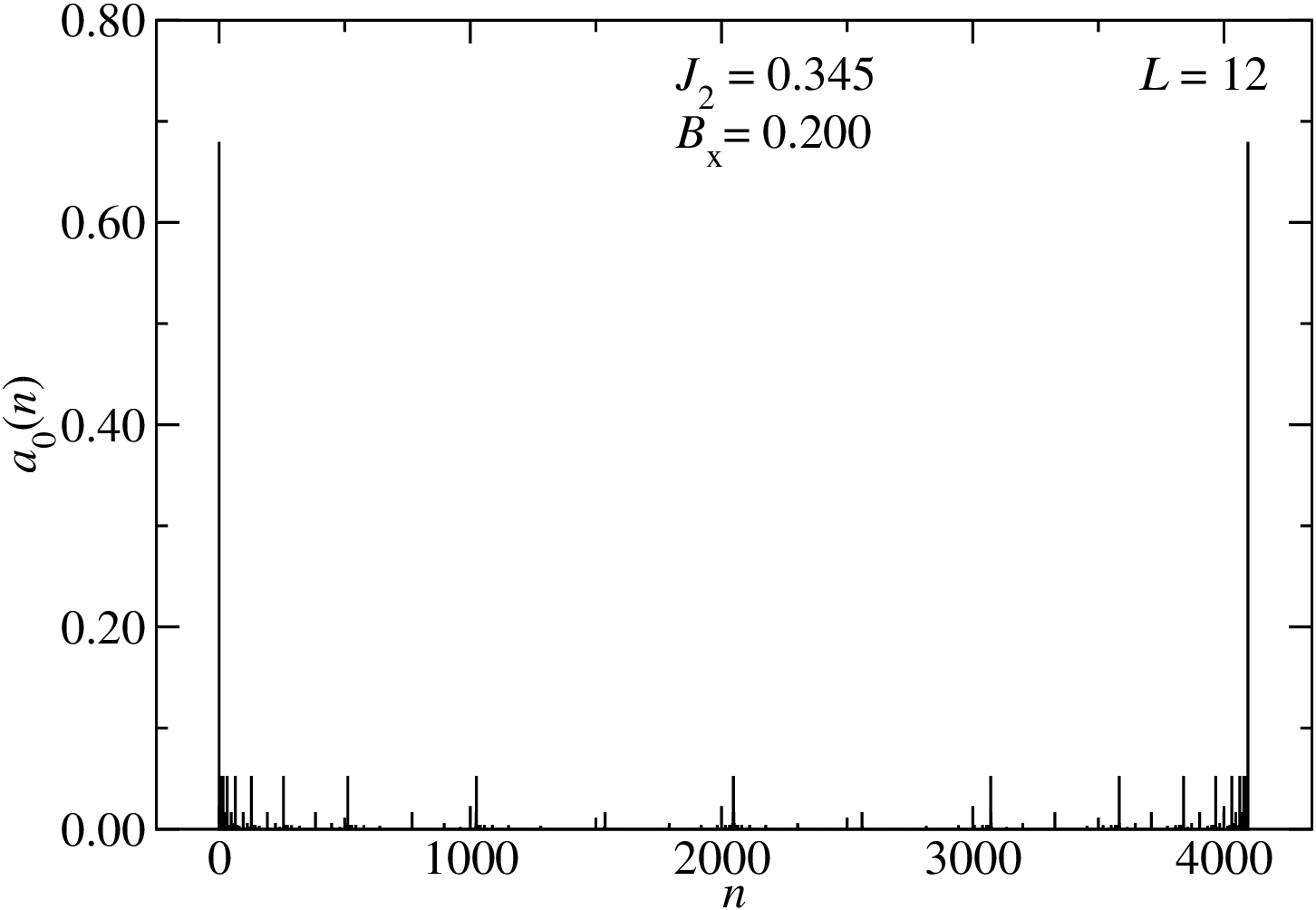}
\setlength{\abovecaptionskip}{0pt}
\setlength{\belowcaptionskip}{10pt}
\caption{Ground-state amplitude versus the basis state 
index $n$ for $(J_2, B_x) = (0.345,0.200)$, within 
the phase $F$ for $L=12$.  
The two largest amplitudes correspond to the ferromagnetic phase. 
The smaller amplitudes are induced by the transverse magnetic field. 
\label{fig:fig8}
}
\end{figure}

\begin{figure}
\includegraphics[width=8.0cm,height= 6.0cm,angle=0]{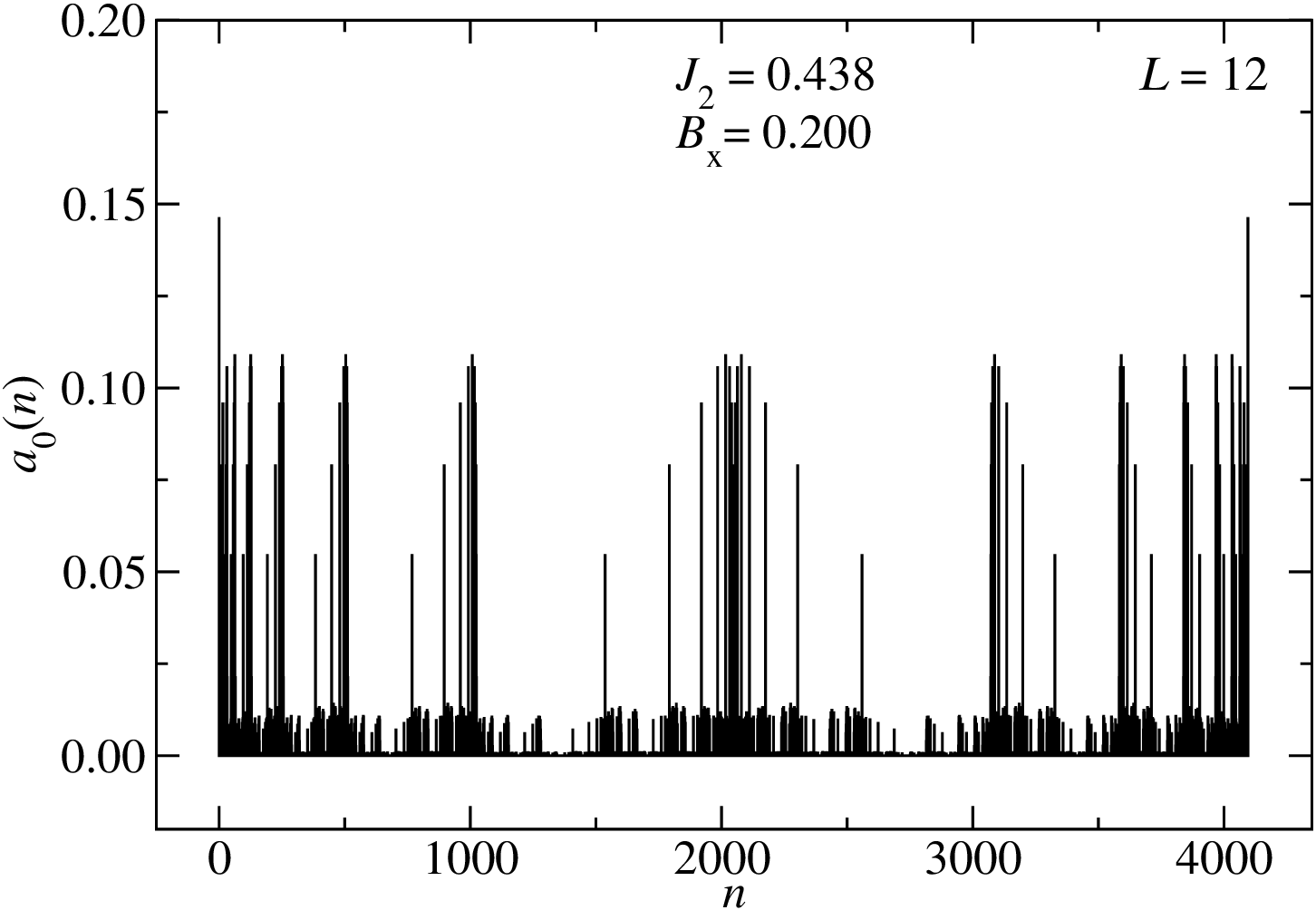}
\setlength{\abovecaptionskip}{0pt}
\setlength{\belowcaptionskip}{70pt}
\caption{Ground-state amplitude for each 
basis state index $n$, with $(J_2, B_x) = (0.438, 0.200)$, 
located inside the phase region $P_2$, for $L = 12$. 
The two largest amplitudes correspond to a ferromagnetic phase. 
The next largest amplitudes are from states with a single-kink  
separating ferromagnetic domains.
\label{fig:fig9}
}
\end{figure}


\begin{figure}
\includegraphics[width=8.0cm, height= 5.9cm, angle=0]{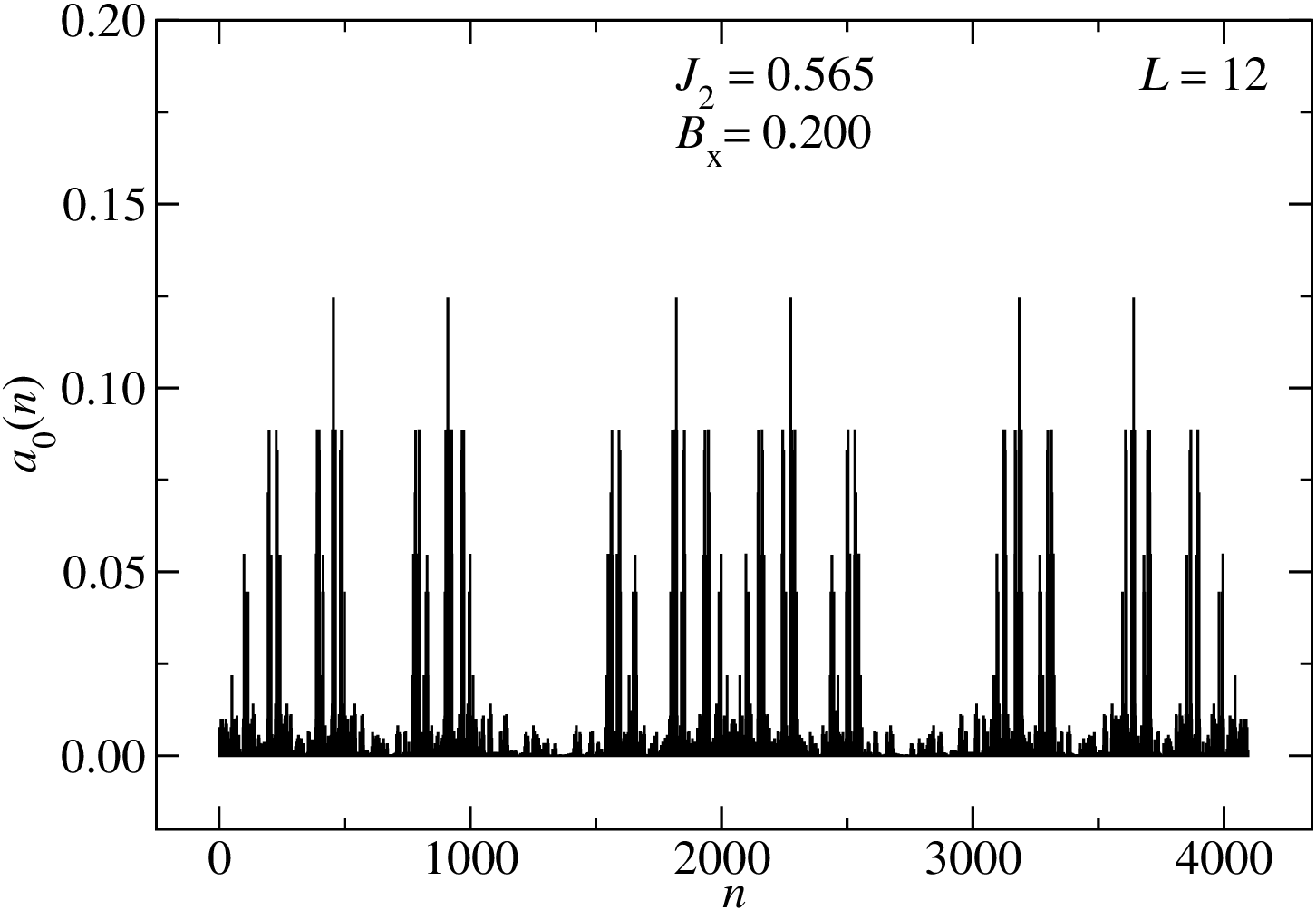}
\setlength{\abovecaptionskip}{0pt}
\setlength{\belowcaptionskip}{20pt}
\end{figure}

\begin{figure}
\includegraphics[width=8.0cm, height= 5.9cm, angle=0]{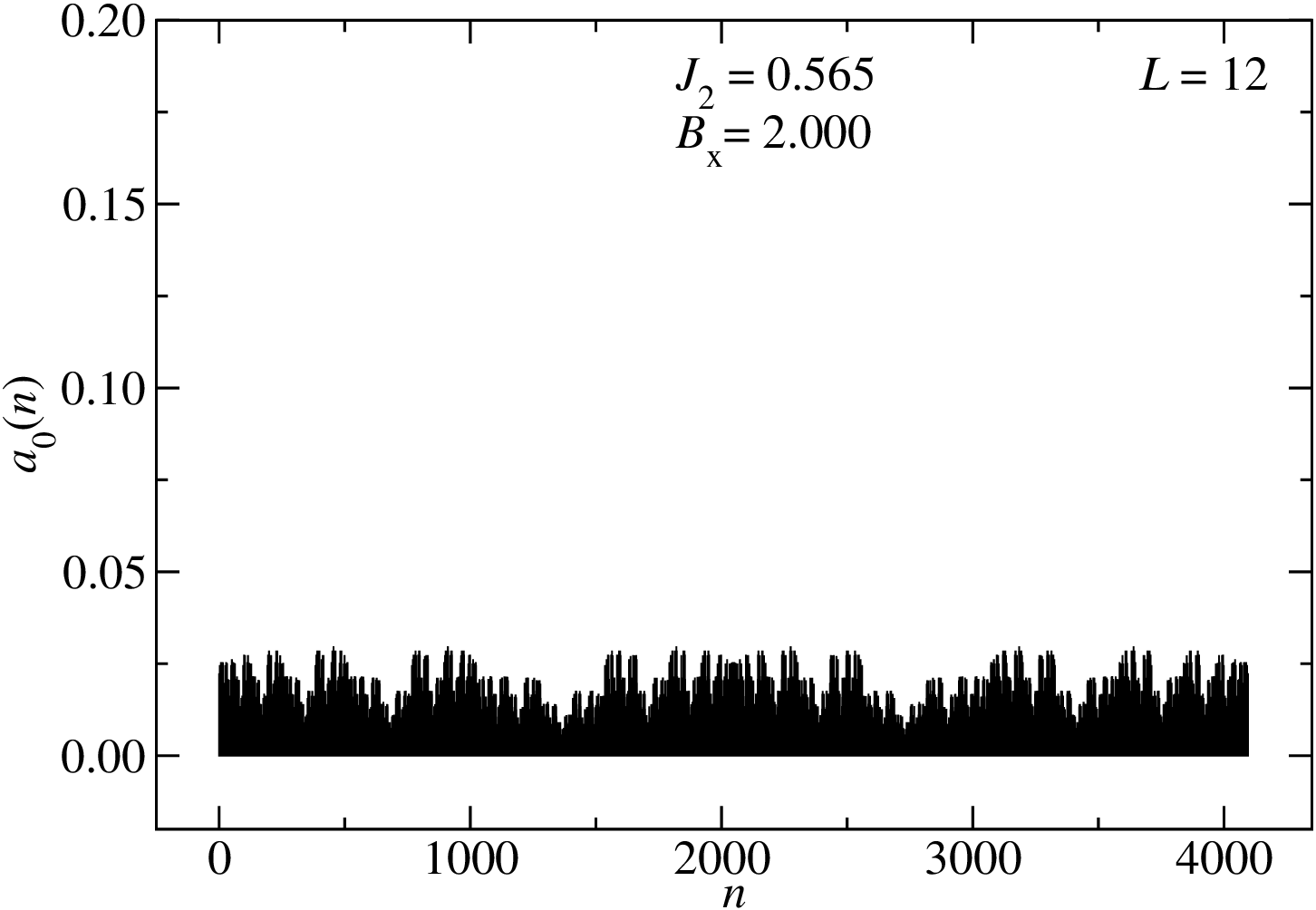}
\setlength{\abovecaptionskip}{0pt}
\setlength{\belowcaptionskip}{20pt}
\end{figure}

\begin{figure}
\includegraphics[width=8.0cm, height= 5.9cm, angle=0]{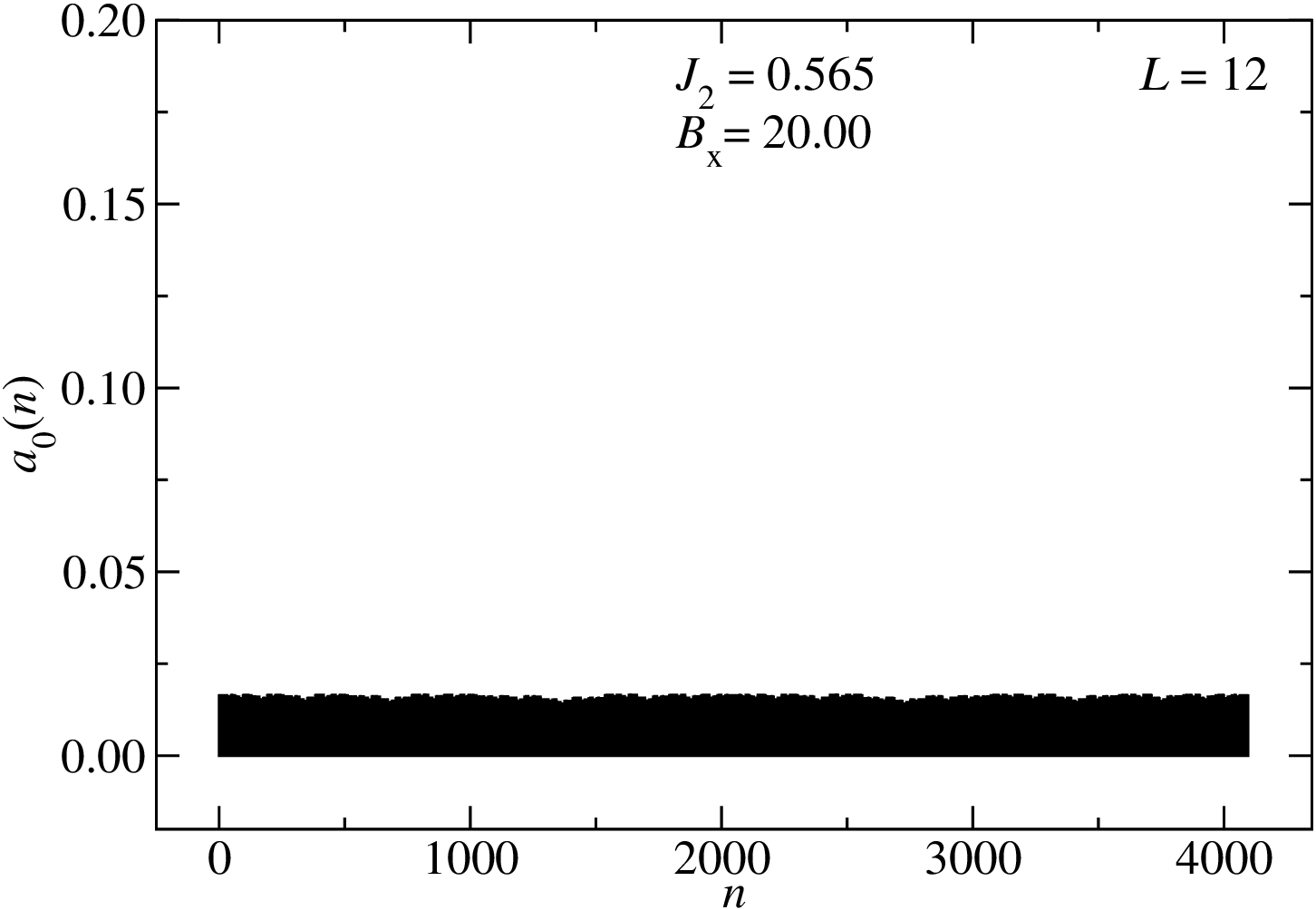}
\setlength{\abovecaptionskip}{0pt}
\setlength{\belowcaptionskip}{20pt}
\caption{Amplitude of the ground-state  against the basis 
state index $n$ for the case $L=12$, $J_2=0.565$, and
different values of $B_x$. 
(Top) $B_x= 0.200$, which lies in the phase
region $P_1$ in Fig.~\ref{fig:fig4}. The six highest amplitudes correspond to a
$\Braket{3,3}$-phase, while the next highest amplitudes
belong to states without sequential order for the spins.
(Middle) $B_x= 2.000$, here there are no noticeable prominent
amplitudes, since the system is already in an induced paramagnetic
state, where the spins are mostly aligned to the transverse field. 
(Bottom) Case $B_x= 20.00$,  now nearly all the spins 
are aligned with the transverse field.
\label{fig:fig10}
}
\end{figure}

\begin{figure}
\includegraphics[width=8.0cm, height= 6.0cm, angle=0]{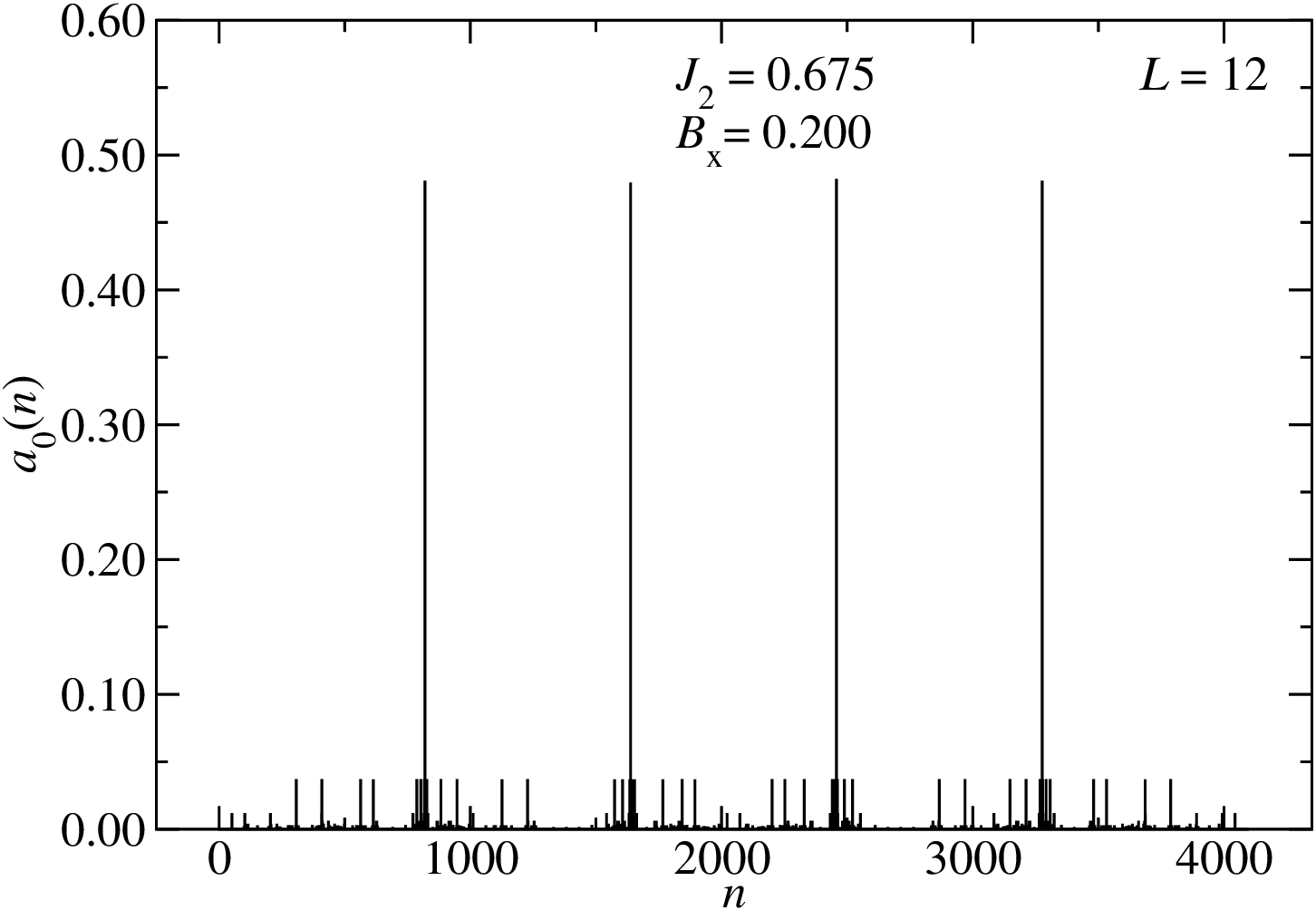}
\setlength{\abovecaptionskip}{0pt}
\setlength{\belowcaptionskip}{30pt}
\caption{Amplitude of the ground-state for each of the basis 
state index $n$ when $(J_2, B_x) = (0.675,0.200)$,
within the phase $\Braket{2,2}$ for $L=12$. 
The four largest amplitudes correspond to the $\Braket{2,2}$-phase. 
The transverse magnetic field is responsible for the appearance 
of the smaller amplitudes. 
\label{fig:fig11}
}
\end{figure}

\begin{figure}
\includegraphics[width=8.0cm, height= 6.0cm, angle=0]{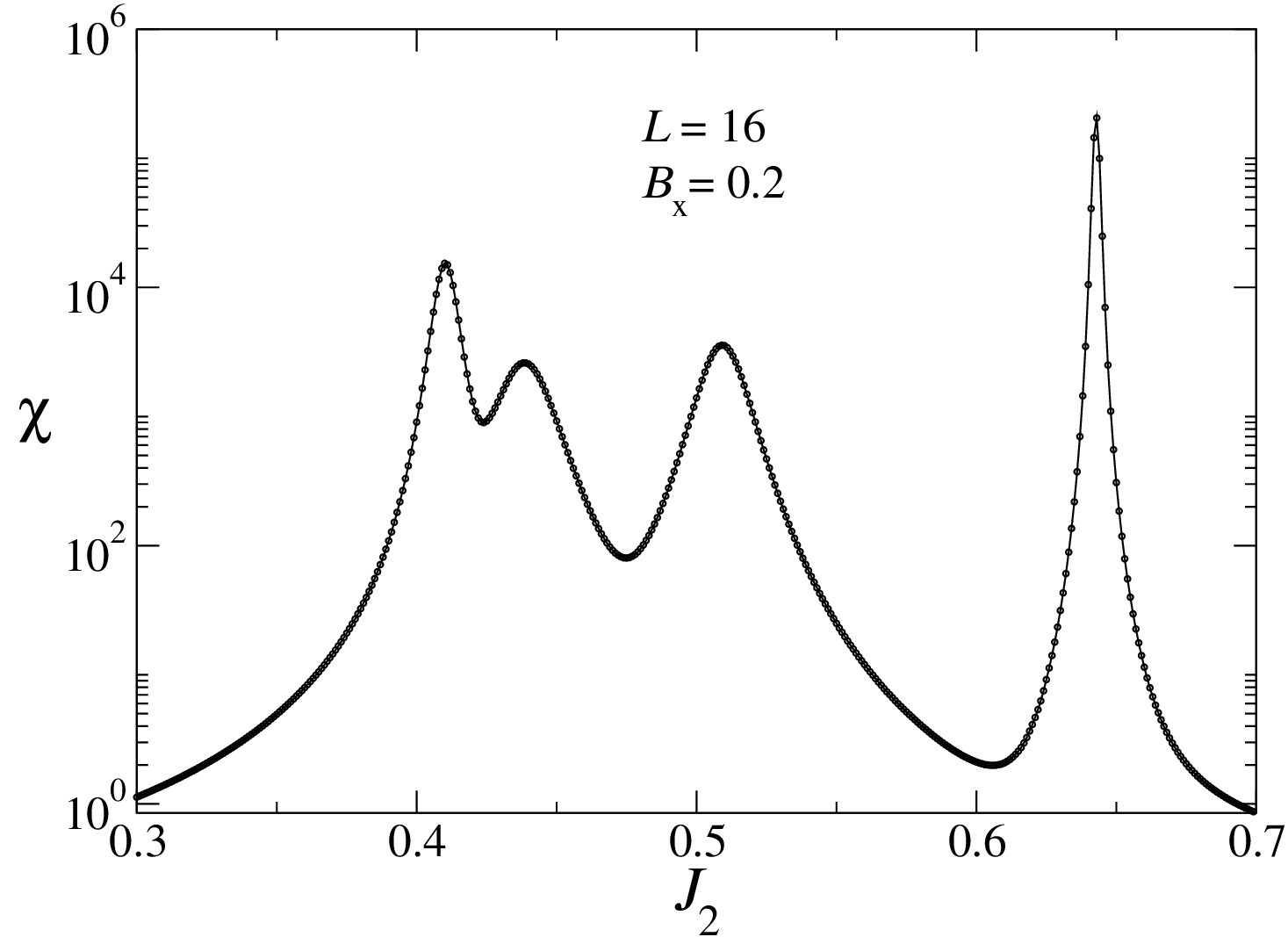}
\setlength{\abovecaptionskip}{0pt}
\setlength{\belowcaptionskip}{20pt}
\caption{Fidelity susceptibility as a function of the next-nearest-neighbor coupling 
for the case $L = 16$. The four peaks shown are centered at the transition points.
Here $B_x = 0.2$.  
 \label{fig:fig12}
}
\end{figure}

\begin{figure}
\includegraphics[width=8.0cm, height= 6.0cm, angle=0]{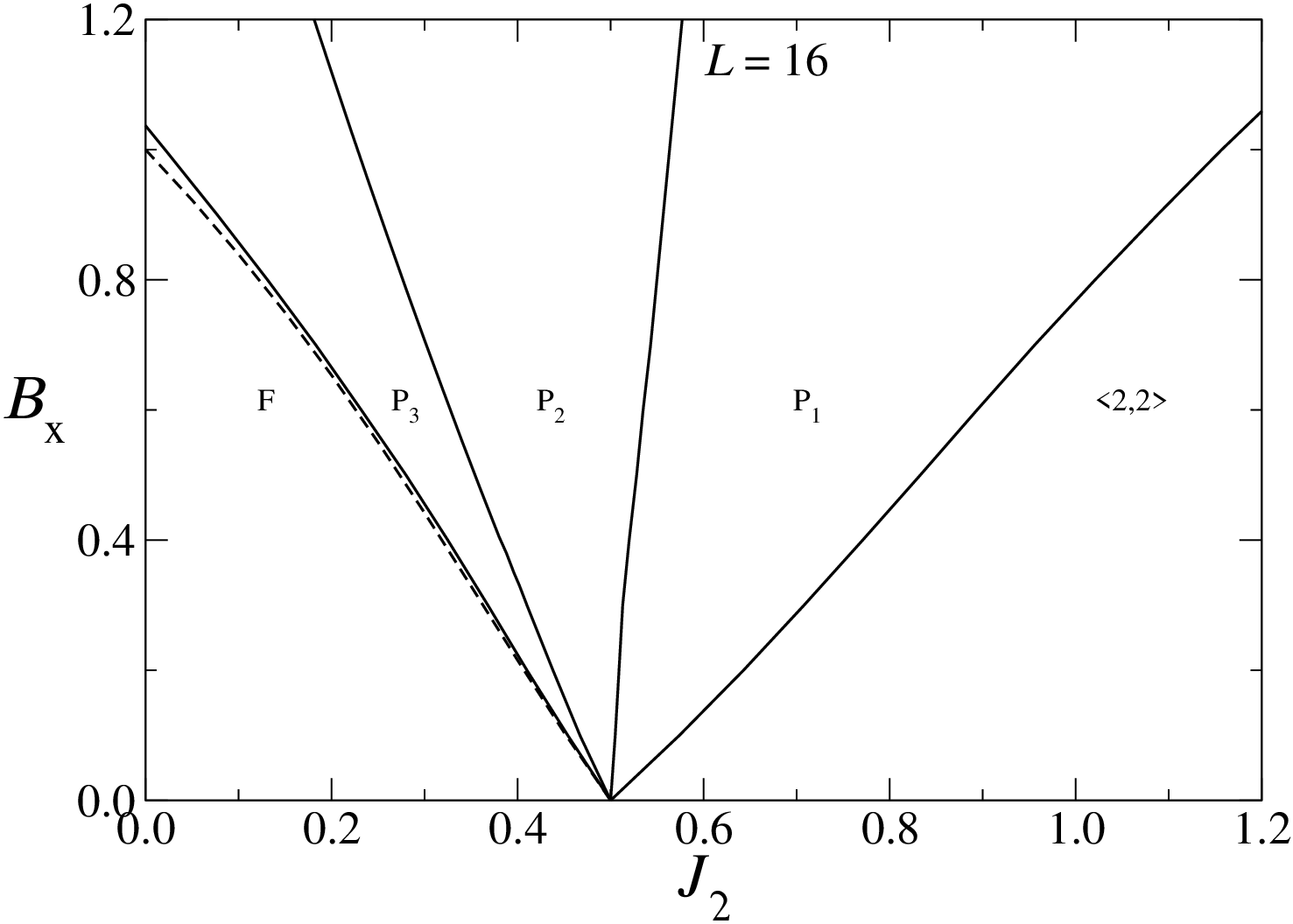}
\setlength{\abovecaptionskip}{0pt}
\setlength{\belowcaptionskip}{30pt}
\caption{Phase diagram in the $(J_2, B_x)$-plane for the case
$L=16$. The figure shows the phase regions:  F,  P$_1$ ,  P$_2$,   P$_3$, and 
$\Braket{2,2}$.
The dashed boundary is the exact Peschel-Emery line.
\label{fig:fig13}
}
\end{figure}

\begin{figure}
\includegraphics[width=8.0cm, height= 6.0cm, angle=0]{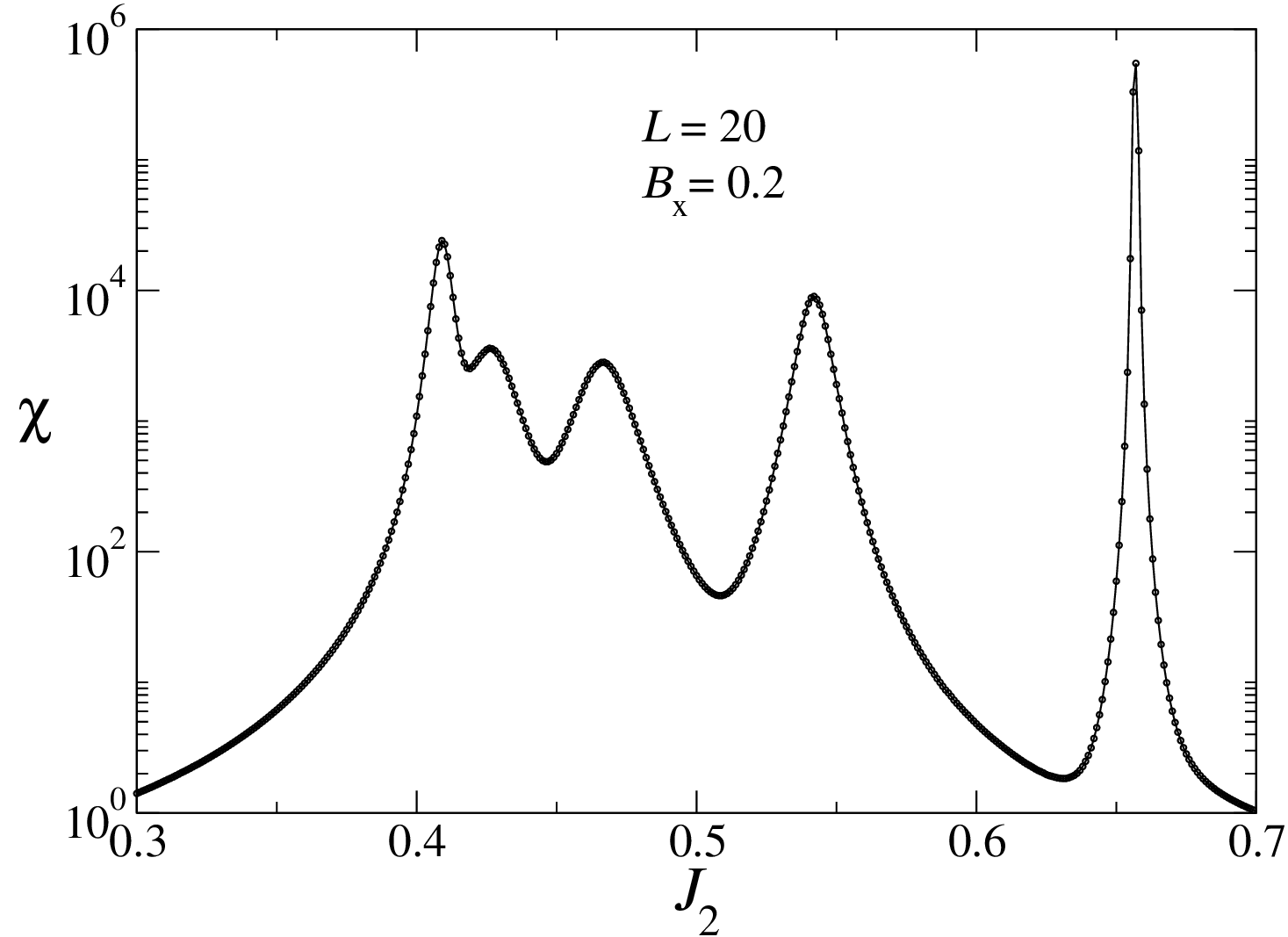}
\setlength{\abovecaptionskip}{0pt}
\setlength{\belowcaptionskip}{20pt}
\caption{Fidelity susceptibility as a function of next-nearest-neighbor coupling
 $J_2$ for $B_x = 0.2$ in the case $L = 20$. The five peaks are centered
  at the transition points. 
 \label{fig:fig14}
}
\end{figure}

\begin{figure}
\includegraphics[width=8.0cm, height= 6.0cm, angle=0]{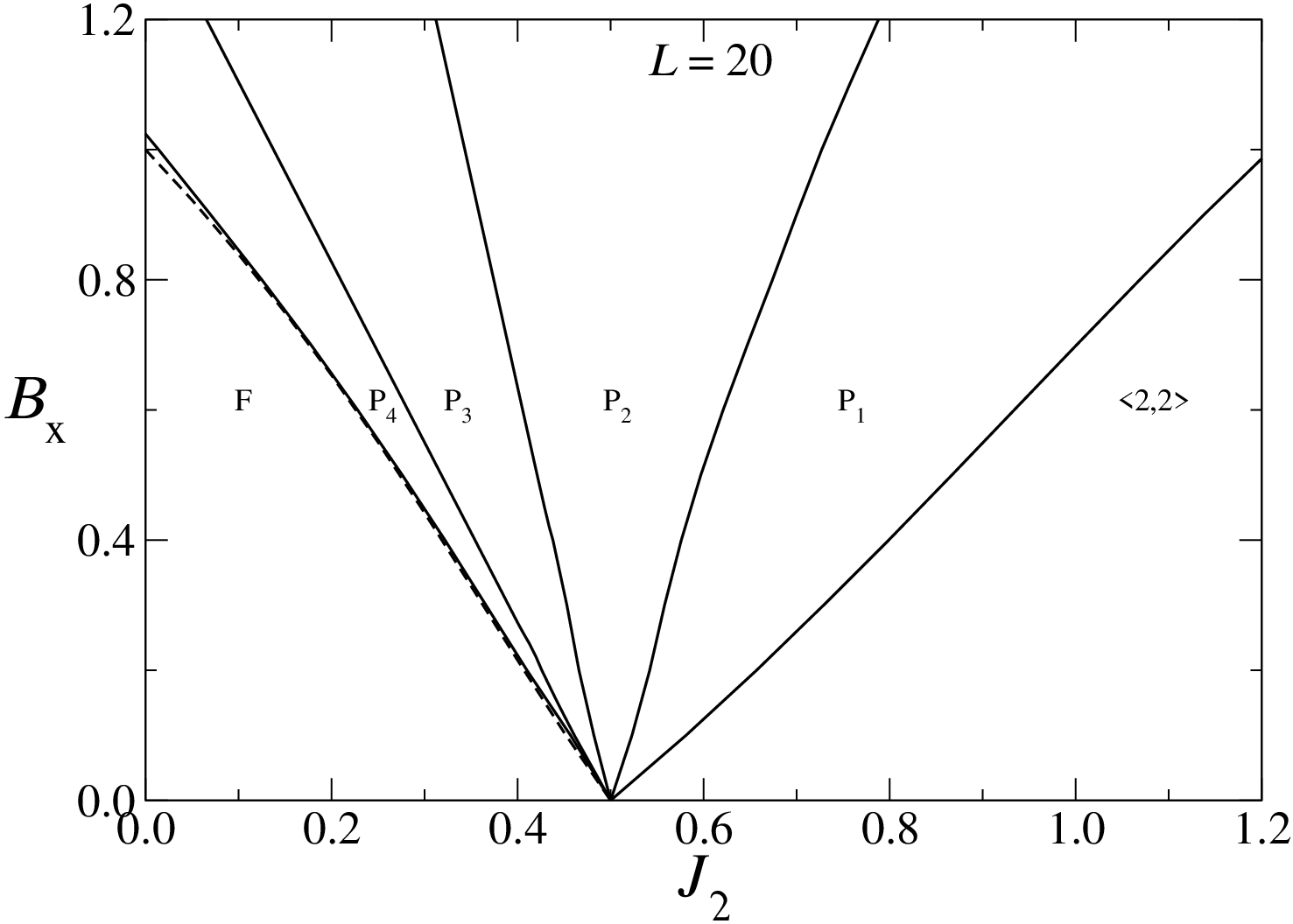}
\setlength{\abovecaptionskip}{0pt}
\setlength{\belowcaptionskip}{30pt}
\caption{Phase diagram in the $(J_2, B_x)$-plane when the system
size is $L=20$.  In addition to the phases $F$ and  $\Braket{2,2}$,
at the left and right of the diagram, respectively, there are four phases in between 
them, namely P$_1$, P$_2$, P$_3$ and P$_4$.
The dashed boundary is the exact Peschel-Emery line.
\label{fig:fig15}
}
\end{figure}

\begin{figure}
\includegraphics[width=8.0cm, height= 6.0cm, angle=0]{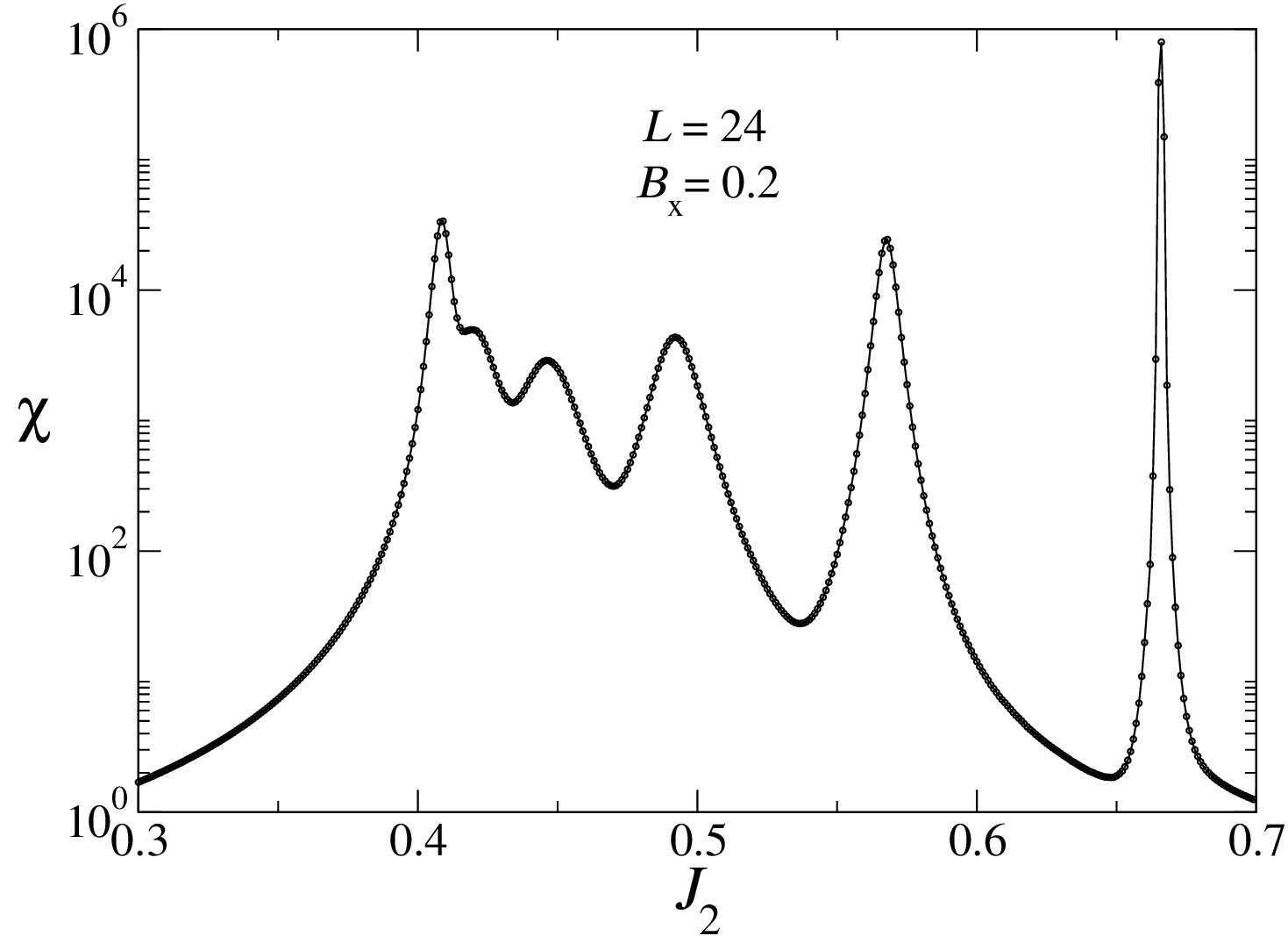}
\setlength{\abovecaptionskip}{0pt}
\setlength{\belowcaptionskip}{20pt}
\caption{Fidelity susceptibility as a function of next-nearest-neighbor coupling
$J_2$  for the case $L = 20$. The six peaks are centered at the transition points. 
 Here $B_x = 0.2$. 
 \label{fig:fig16}
}
\end{figure}

\begin{figure}
\includegraphics[width=8.0cm, height= 6.0cm, angle=0]{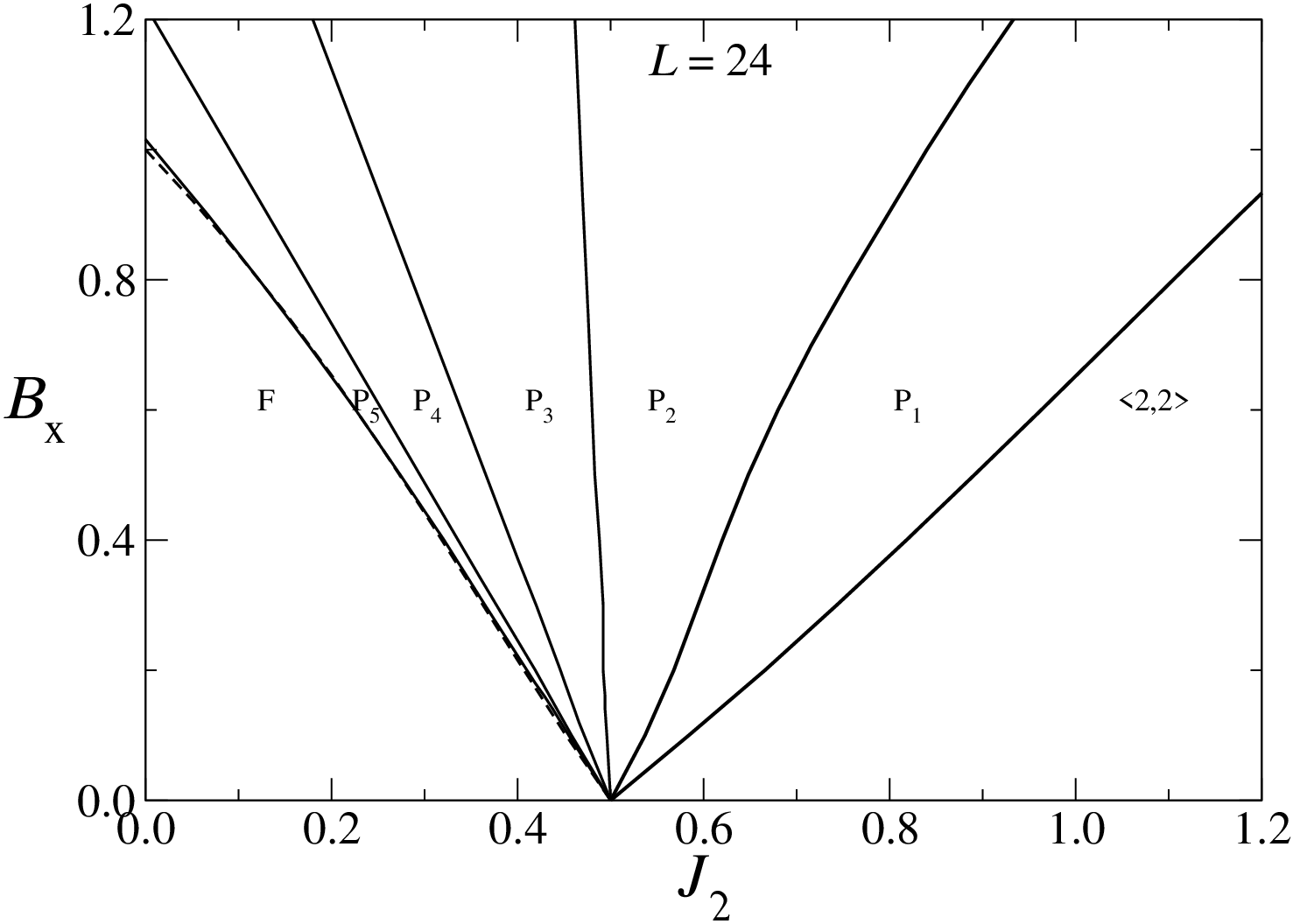}
\setlength{\abovecaptionskip}{0pt}
\setlength{\belowcaptionskip}{40pt}
\caption{Phase diagram in the $(J_2, B_x)$-plane when the system
size is $L=24$.  In addition to the phases $F$ and  $\Braket{2,2}$,
at the left and right of the diagram, there are five phases in between 
them, namely P$_1$, P$_2$, P$_3$, P$_4$, and P$_5$.
The dashed boundary is the exact Peschel-Emery line, which lies very
close to transition line between $F$ and $P_5$.
\label{fig:fig17}
}
\end{figure}

\begin{figure}
\includegraphics[width=8.0cm, height= 6.0cm, angle=0]{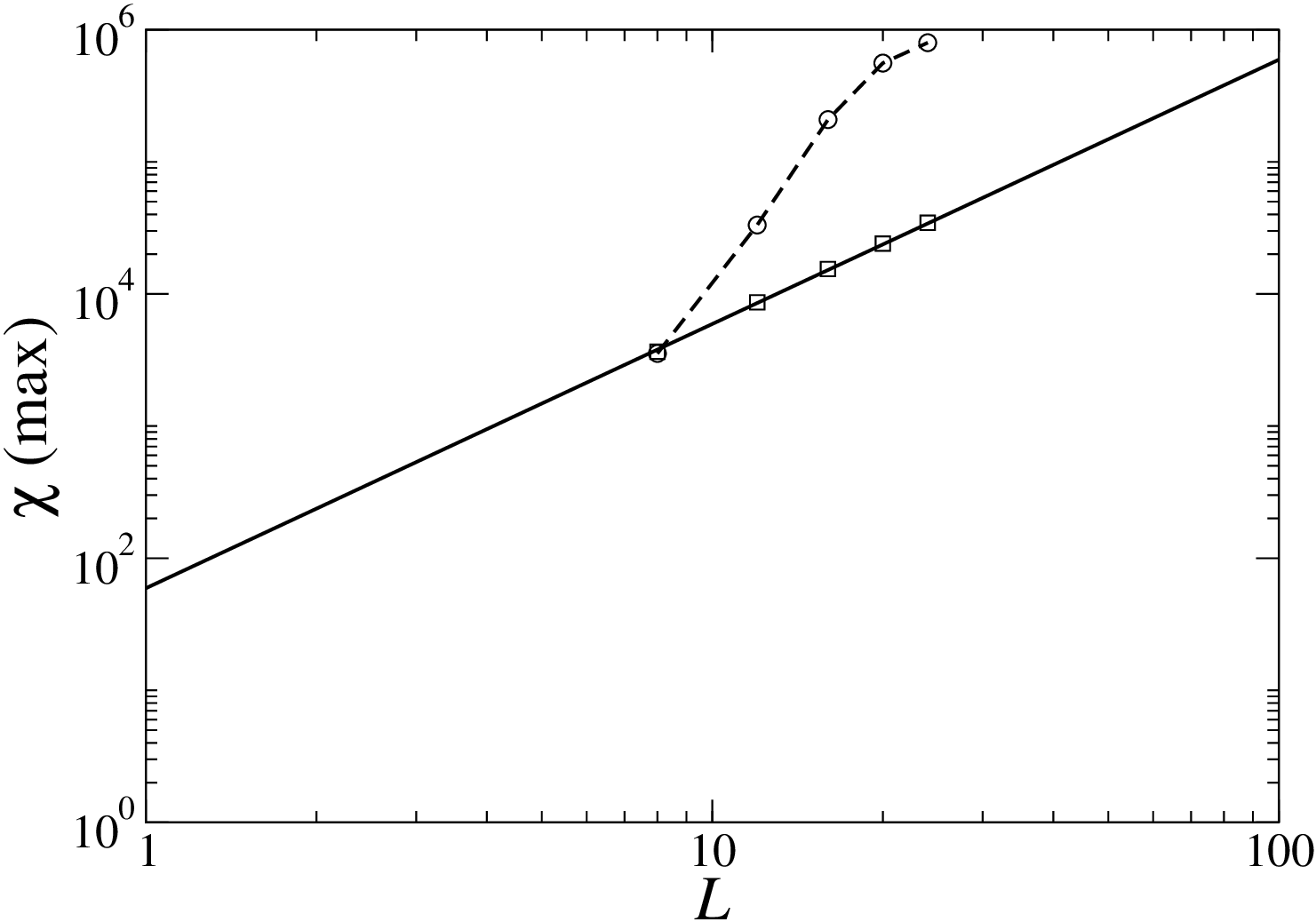}
\setlength{\abovecaptionskip}{0pt}
\setlength{\belowcaptionskip}{20pt}
\caption{Fidelity susceptibility at criticality as a function of the lattice size $L$
for two different transition lines.
Squares are for the transition line bordering the ferromagnetic phase (Peschel-Emery line)
while circles are for the antiphase.
 \label{fig:fig18}
}
\end{figure}

In order to identify the nature of the quantum phase, 
we examined how  the ground-state 
eigenvectors are written in terms of a complete 
set of appropriate basis vectors.
To find the eigenstates and corresponding 
eigenvalues of the system we needed to choose 
a complete set of orthogonal basis vectors and 
write the Hamiltonian in matrix form using this basis set.
The eigenvalues and eigenstates are found by exact numerical
diagonalization. 
A convenient basis consists of the tensor product 
of $L$ eigenstates of the z-component of the 
local spin-operator acting on each site. 
Denoting the eigenstates by $|s>_i$, 
where  $s=1$,  is the eigenstate label of the 
operator $\sigma^z_{i}$ for an up-spin 
and $s = 0$ for the a down-spin at site $i$. 
A generic basis eigenstate for the full system 
with $L$ spins can be written as $|n> = \prod_i^L\, |s>_i$,  
where $n$ labels the basis state and has 
the values $n = 0, 1,..., N-1$, and
where $N = 2^L$  represents the dimension 
of the Hilbert space. 
The basis index $n$, if written in binary notation, 
can also be used to specify the configuration 
of the spins forming that basis. 
That is, when $n$ is written in binary notation, 
the position and value of a bit will indicate 
whether the spin at that position (site) is up (1) 
or down (0). 
For instance, for a chain of 12 spins
 the state $|$1755$>$ in binary notation is written as 
$|$011011011011$>$, which represents 
a periodic configuration with one down-spin (0) followed by two 
up-spins (11). 
In this notation, an arbitrary eigenstate of the Hamiltonian 
may be cast as 
\begin{equation}
|\phi_{\alpha}> = \sum_{n=0}^{N-1}a_{\alpha}(n)|n>,
\label{eq:phi}
\end{equation}
where  $\alpha = 0, ...,N-1$, labels the quantum states, 
with  $\alpha = 0$ assigned to the ground-state.
Since the matrix Hamiltonian  is  real and symmetric, 
the coefficients $a_{\alpha}(n)$ are real. 
As a result, the quantum state $|\phi_{\alpha}>$ can be 
visualized in a single graph by plotting 
$a_{\alpha}(n)$ as a function of the quantum state index $n$. 
The graph will completely identify the spatial distribution 
of spins in the quantum state \cite{Bon06,Boe14, Bon14}.

\section{\label{sec:IV} Results} 

 In the following we present our results for system sizes
$L= 8$, $12$, $16$, $20$, and $24$.  We chose
those sizes in order to avoid the effects of frustration and preserve 
the symmetry of the $\Braket{2,2}$ antiphase, which has periodicity of
4 lattice spacings.  Still we are able to draw reliable conclusions
as well as predictions about the quantum model in
the thermodynamic limit.

Let us consider first the case $L=8$. 
Figure~\ref{fig:fig1} shows the fidelity susceptibility plotted against the 
next-nearest-neighbor interaction $J_2$ 
for a fixed transverse field $B_x = 0.2$. 
The two peaks in the graph give 
the locations of the critical points where quantum
phase transitions occur. 
By calculating the susceptibility for several  values of $B_x$ 
and $J_2$, we obtain the phase diagram shown in Fig.~\ref{fig:fig2}. 
There, we readily identify three distinct phases for
low magnetic fields.
The  region farthest to the left (F) is  
ferromagnetic, while the middle (P$_1$) has a
modulated phase, and  the region farthest to the
right has the antiphase  ($\Braket{2,2}$).
The transition line bordering the ferromagnetic 
phase is close to the exact Peschel-Emery 
line~\cite{Pes81}.  As we shall see, for larger system
sizes we obtain results which closer to that line.
Notice that all the phase boundary lines
meet at $(J_2,B_x)=(0.5,0.0)$,  
the known multicritical point.
Finally, for large enough magnetic fields,
the modulated phase becomes paramagnetic.
Such a feature does not appear in the phase
diagram shown, which covers relatively low magnetic
fields, where lies the interesting physics.
That is also true for all the following phase diagrams
below, which are valid at the low field region, where we 
are concerned with the onset and further evolution 
of modulated phases as the system sizes increases.

Consider now $L=12$. 
Figure~\ref{fig:fig3} shows the fidelity susceptibility versus
$J_2$, for  $B_x = 0.2$. 
The three peaks on the graph give 
the locations  where the 
phase transitions occur. 
Proceeding in a similar way for various values
of $B_x$ we determine the phase diagram,
which is shown in Fig.~\ref{fig:fig4}.
Alternately, by keeping $J_2$ fixed and sweeping
with $B_x$ we obtain the same phase diagram.
As an example of this we present Figs.~\ref{fig:fig5}, \ref{fig:fig6}, and \ref{fig:fig7},
which shows the susceptibilities along $B_x$.
The peaks are at the same locations as those obtained
earlier with $J_2$ sweeps.
As can be seen, there appears an additional phase boundary line,
as compared to the case $L=8$.
There is a modulated phase in the region $P_2$,
and a floating phase $P_1$.  These phases
are separated by the boundary line that
 meets at the multicritical point. 
For very large fields $B_x$ we expect the system to be
paramagnetic.  The ferromagnetic 
and antiphase regions remain basically
the same, apart from a slight shift in their 
borders, due to finite size effects.
The boundary line between the ferromagnetic
and its neighboring modulated phase is now closer to 
the Peschel-Emery line than that of the case $L=8$.

The spin configurations in each of the phases
can be inferred from a plot of the amplitudes
$a_0(n)$ of the ground-state eigenvector
versus the basis index $n$ for a point
deep within a given phase.  For instance,
consider the point in the phase diagram 
$(J_2,B_x)=(0.345,0.200)$, which is in the F-phase.
Figure~\ref{fig:fig8} shows $a_0(n)$ {\em vs} $n$
for that point.  The two largest contributions to
the ground-state correspond to the ferromagnetic  
spin configurations, $n=0$ and $n=4095$, which have
 binary representations $\Ket{000000000000}$ and 
 $\Ket{111111111111}$, respectively. 
The other basis states with smaller   amplitudes are
induced by the transverse magnetic field.  
Those amplitudes increase with  $B_x$.
Consider now $(J_2,B_x)=(0.438,0.200)$, which lies in
the region $P_2$ of Fig.~\ref{fig:fig4}. 
The amplitudes of the ground-state basis vectors
are depicted in Fig.~\ref{fig:fig9}.  The largest contributions
come from ferromagnetic orderings, while 
the second largest amplitudes are from
the basis state $\Ket{000000111111}$
and its cyclic permutations of the spins. The third largest
amplitudes are very close to the second.
They come from the states $\Ket{000000011111}$, 
$\Ket{111111100000}$, and all the others
were obtained by their cyclic relatives.
The boundary line separating the F-phase from the neighboring
modulated phase starts out at the multiphase point $(J_2,B_x)=(0.5,0.0)$
and ends close to the Pfeuty transition point $(J_2,B_x)=(0.0,1.0)$.

We find that as the transverse field becomes sufficiently large the
system enters a paramagnetic phase, where the spins tend
to point in the same direction as the field.  That is a general feature of the model.
No matter which phase the system is in when $B_x$ is small, eventually it
will become paramagnetic as the field increases.  We do not
find any evidence of a sharp transition to paramagnetism.
It seems that paramagnetism is achieved through a crossover
mechanism, so that no transition line is observed.
Fig.~\ref{fig:fig10} shows the ground-state eigenvector
amplitudes for 3 cases: $B_x=0.200$, $2.000$, and $20.00$.
The figures were obtained for $L=12$ and $J_2=0.565$,
but similar behavior is expected for any other set of
parameters $L$ and $J_2$. The top figure ($B_x=0.2$)
shows 6 largest amplitudes that correspond to
that basis vectors containing periodic sequences of 3 up-
followed by 3 down-spins.  The next largest amplitudes
stem from spin arrangements not periodic.
As the field becomes sufficiently large, the amplitudes
for the ordered phase disappear, while all the other
amplitudes becomes larger, as can be seen
in the middle figure of Fig.~\ref{fig:fig10}.   There, most
of the spins are equally likely to align themselves with
the transverse field. Finally, for very large fields (e. g., $B_x=20.00$),
 nearly all the spins align themselves with the field, resulting
in a more evenly distributed amplitudes of the basis
vectors.  Clearly the system is in an induced paramagnetic
phase.
As we shall see later, when we consider larger lattices,
nonperiodic configurations will dominate the low-$B_x$
phase. That amounts to the so-called floating phase.
In that phase there is not any periodic spin order
commensurate with the underlying lattice.

Finally, the ground-state of the rightmost phase 
in Fig.~\ref{fig:fig4} is dominated by four amplitudes corresponding 
to the $\Braket{2,2}$-phase. 
 The dependence of the amplitudes with the state index  
 for $(J_2,B_x)=(0.675,0.200)$ in that phase, 
is depicted in Fig.~\ref{fig:fig11}. 
 Again, small amplitudes are due to the 
 transverse magnetic field and, as in the other cases, 
and they get larger as $B_x$ increases.

Both the F-phase and the $\Braket{2,2}$-phase are present 
in all the cases we considered ($B_x \le 1.2$),
for all lattice sizes $L$.  
They are expected to be present 
in the thermodynamic limit. This is in agreement with the results 
found by other methods~\cite{Gui02,Bec06,Bec07,Ari91,Cha07,Nag11}.
However, as we consider larger lattices, 
other modulated phases  appear in between the ferromagnetic 
and the floating phase.  It should be noted that all the transition 
lines start out at the multiphase point and then spread 
outwards as $B_x$ increases.  For sufficiently large $B_x$
the phase is expected to be paramagnetic.

Let us consider now the model with size $L=16$.
Figure~\ref{fig:fig12} shows the fidelity susceptibility as a function 
of  $J_2$, for $B_x=0.2$.  
The susceptibility exhibits four peaks, thus indicating 
five distinct phases. 
Again, by numerically varying  $B_x$ and $J_2$, 
we obtained the phase diagram for the system,  
depicted in Fig.~\ref{fig:fig13}.
At the two far sides of the diagram
we obtained the F- and $\Braket{2,2}$-phases, as in the
previous case. 
The positions of the boundaries of the F- and $\Braket{2,2}$-phases
with their neighboring phases 
are weakly dependent on the system size, 
especially the boundary of the F-phase.  
The slope of the boundary line of the  $\Braket{2,2}$-phase
for $L=16$ diminishes a little as compared with the previous case $L=12$.
We find an additional modulated phase,
which is dominated by states with the 
ordered pattern  $\Braket{4,4}$.
There appears to be other contributions to the ground-state
of much smaller weights which are not ordered, but which will
increase with the applied field $B_x$. 
Again, all the transition lines
start at the multicritical point.

For $L=20$ and $B_x= 0.2$ the fidelity susceptibility 
shows 5 peaks, as seen in Fig.~\ref{fig:fig14}. 
The plot indicates the existence of five phase transitions 
for this lattice size. The phase diagram $J_2$--$B_x$ 
is shown in Fig.~\ref{fig:fig15}. We observe that another
 modulated phase has appeared. 
Now, in addition to the ferromagnetic, floating, 
and  $\Braket{2,2}$-antiphase, the system now has  
three modulated phases. The floating phase P$_1$ for this lattice size is 
dominated by the orderings $\Braket{3,2}$ and  $\Braket{2,3}$.
Again, the modulated phases eventually  become 
paramagnetic for large enough transverse fields.

For larger system sizes, we observe a pattern that 
allows us to make inferences about the phases of the system in the 
thermodynamic limit. Due to computer limitations,
the largest system studied is $L=24$.
Figure~\ref{fig:fig16} shows the fidelity susceptibility as a function of
 $J_2$, for  $B_x=0.2$. There are 6 peaks, indicating an
equal number of phase transitions. 
The phase diagram is shown in Fig.~\ref{fig:fig17}.  
We now identify 4  modulated phases in the figure,
P$_2$, P$_3$, P$_4$, and P$_5$, in addition to
the floating P$_1$, ferromagnetic F, and the
$\Braket{2,2}$ phases. The paramagnetic phase
only occurs for high $B_x$, where the phases
lose their characteristics as the spins tend to
align with the transverse field.
The modulated phases are characterized
by several periodicities, among them
$\Braket{4,4}$ for P$_3$,
and $\Braket{3,3}$ for P$_2$.
The floating phase P$_1$ is now dominated
by configurations which do not exhibit
any periodicity within the system size.
No particular ordering seems to take place
as $L$ increases, hence no commensurate 
order emerges in the floating phase.

As the system size increases, more modulated phases
appear.
For sufficiently large transverse magnetic fields one expects
the system to become paramagnetic.
 The origin of the modulated phases follows from the degeneracy 
 of the ground-state at $J_2= 0.5$ and $B_x=0.0$. 
 There, the ground-state is highly degenerate, with the number 
 of configurations exponentially increasing with the size of the system, as mentioned before. 
 The transverse magnetic field lifts the degeneracies, thus separating the phases.  
 At finite sizes, some of the phases become visible.
 As one consider larger systems, more
of those  phases appear.
 The ferromagnetic as well as the $\Braket{2,2}$ phases should be obviously
 present for any system size in the cases $J_2 <0.5$ and $J_2>0.5$, 
 respectively, since they are energetically favorable in those
 situations.
  Our numerical analysis was done with a
 maximum of 24 spins due to computer limitations.
 Yet, we can expect that as the number
 of spins increases there will appear more and
 more distinct modulated phases.
 We predict that at the thermodynamic limit
 there will be a (denumerable) infinite number
 of modulated phases. 
 
 At criticality the fidelity susceptibilty shows power-law behavior
 with the lattice size, indicating that the transition is of second-order;
 otherwise it is of first-order~\cite{You11}.  For instance, for the
 transition line closest to the feromagnetic phase we observe a power-law
 behavior, which is shown in Fig.~\ref{fig:fig18}.
 The solid line is the numerical fit $\chi = 59.2L^2$. 
It seems that all the transition lines between modulated
phases are of second-order.  In particular, the transition
between the modulated phase P$_2$ and the floating phase (P$_1$) 
seems to be of second-order, contrary to the claims that it is of BKT type.
Finally, the transition line separating the floating and 
$\Braket{2,2}$ antiphase is of first-order, since the behavior of the susceptibility 
deviates from  power-law, as can be seen in Fig.~\ref{fig:fig18}.

The scaling behavior of the fidelity susceptibility in the vicinity 
of a quantum critical point has been found to be \cite{Sch09,Alb10}:
\begin{equation}
\chi(\lambda_c) \sim L^{2/\nu}.
\label{eq:exponent}
\end{equation}
where  $\nu$ is the critical exponent describing the divergence 
of the correlation function.  For the case of the transition line 
closest to the ferromagnetic phase (see Fig.~\ref{fig:fig18}), 
the behavior of the fidelity susceptibiliy at criticality is quadratic 
implying that $\nu =1$. Hence, in this region the model is in 
the same universality class as the transverse Ising model. 

\section{\label{sec:V} Summary and Conclusions}

We have studied the ground-state 
properties of the one-dimensional ANNNI model 
in a transverse magnetic field. 
The phase diagrams in the ($J_2$, $B_x$) plane were 
obtained using the quantum fidelity method 
for several lattice sizes.  
A new picture emerged that is distinct from previously 
reported results.  
In addition to the known phases, namely,
ferromagnetic,  floating, and the $\Braket{2,2}$ phase, 
it seems that there will be an infinite number of modulated phases
of spin sequences commensurate with the underlying 
lattice in the thermodynamic limit.  We do not find paramagnetism for
small values of the applied field.  Paramagnetism is expected
 to occur at sufficiently high fields, not shown in our phase
diagrams.
The transitions between the modulated phases
seem to be of second-order. On the other hand, 
the transition between the floating and $\Braket{2,2}$ phase
appears to be of first-order.

\vskip1cm
\centerline{\bf ACKNOWLEDGEMENTS}
\vskip0.5cm
We thank C. Warner for critical reading of the manuscript.
We also thank FAPERJ, CNPq and PROPPI (UFF)  for financial support.
O.F.A.B. acknowledges support from the Murdoch
College of Science Research Program and a grant from the 
Research Corporation through the Cottrell College Science 
Award No.~CC5737.


\begin{thebibliography}{}

\bibitem{Sac99} S.~Sachdev, {\em Quantum Phase Transitions} (Cambridge 
University Press, Cambridge, England, 1999).

\bibitem{Gen63} P. G.~de Gennes, Solid State Comm. {\bf 1}, 132 (1963).

\bibitem{Bit96} D.~Bitko, T.F.~Rosenbaum, and G.~Aeppli, Phys. Rev. Lett. 
{\bf 77}, 940 (1996).

\bibitem{Pfe70} P.~Pfeuty, Ann. Phys. (N.Y.) {\bf 57}, 79 (1970).

\bibitem{Sim11} J.~Simon, W. S.~Bakr, R.~Ma, M. E.~Tai, P. M.~Preiss, 
and M.~Greiner, Nature (London) {\bf 472}, 307 (2011).

\bibitem{You75} A. P.~Young, J. Phys. C  {\bf 8}, L309 (1975).

\bibitem{Her76} J. A.~Hertz, Phys. Rev. B {\bf 14}, 1165 (1976).

\bibitem{Suz76} M.~Suzuki, Prog. Theor. Phys. {\bf 56}, 1454 (1976).

\bibitem{Ruj81} P. Rujan, Phys. Rev. B 24, 6620 (1981).

\bibitem{Bar81} M.N.~Barber and P.M.~Duxbury, J. Phys. A {\bf 14}, L251 (1981).

\bibitem{Bar82} M.N.~Barber and P.M.~Duxbury, J. Stat. Phys. {\bf 29}, 427 (1982).

\bibitem{Suz13} S.~Suzuki, J.~Inoue, and D.K.~Chakrabarty, {\em Quantum Ising Phases and Transitions in Transverse Ising Models}, Lecture Notes in Physics, 2nd ed., Heidelberg, Springer, 2013.

\bibitem{Dut15} A.~Dutta, G.~Aeppli, B.K.~Chakrabarti, U.~Divakaran, T.F.~Rosenbaum, and D.~Sen, 
in {\em Quantum Phase Transitions in Transverse Field Spin Models: From Statistical Physics to Quantum Information}, Cambridge University Press, Delhi, (2015).

\bibitem{Ari91} C. M.~Arizmendi, A. H.~Rizzo, L. N.~Epele, and C. A.~Garcia, 
Z. Phys. B: Condens. Matter {\bf 83}, 273 (1991).

\bibitem{Sen92} P.~Sen, S.~Chakraborty, S.~Dasgupta, and B.K.~Chakrabarti, 
Z. Phys. B: Condens. Matter {\bf 88}, 333 (1992).

\bibitem{Rie96}H.~Rieger and G.~Uimin, Z. Phys. B: Condens. Matter 
{\bf 101}, 597 (1996).

\bibitem{Sen97} P.~Sen, Phys. Rev. B, {\bf 55}, 11367 (1997).

\bibitem{Gui02} Paulo R.~Colares Guimar\~aes, J. A.~Plascak, F. C.~S\'a Barreto, 
and J.~Florencio, Phys. Rev. B, {\bf 66}, 064413 (2002).

\bibitem{Dut03}A.~Dutta and D.~Sen, Phys. Rev. B, {\bf 67}, 094435 (2003).

\bibitem{Bec06} M.~Beccaria, M.~Campostrini, and A.~Feo, Phys. Rev. B 
{\bf 73}, 052402 (2006).

\bibitem{Bec07} M.~Beccaria, M.~Campostrini, and A.~Feo, Phys. Rev. B {\bf 76}, 094410 (2007).

\bibitem{Cha07} A. K.~Chandra and S.~Dasgupta, Phys. Rev. E 75, 021105 (2007).

\bibitem{Nag11} A.~Nagy, New J. Phys. {\bf 13} 023015 (2011).

\bibitem{Fis80} M. E.~Fisher and  W.~Selke, Phys. Rev. Lett. 44, 1502 (1980).

\bibitem{Red81} S.~Redner, J. Stat. Phys. {\bf 25}, 15 (1981).

\bibitem{Sel88}W.~Selke, Phys. Rep. {\bf 170}, 213 (1988).

\bibitem{Pes81} I.~Peschel and V.J.~Emery, Z. Phys. B: Condens. Matter {\bf 43}, 241 (1981).

\bibitem{And67} P. W.~Anderson, Phys. Rev. Lett. {\bf 18}, 1049 (1967).

\bibitem{Zan06} P.~Zanardi and N.~Paunkovi\'c, Phys. Rev. E {\bf74}, 
031123 (2006).

\bibitem{Ben92} C. H. Bennett, G. Brassard, and N. D. Mermin, Phys. Rev. Lett. 
{\bf 68}, 557 (1992).

\bibitem{Aba08} D. F. Abasto, A. Hamma, and P. Zanardi, Phys. Rev. A 78, 
010301 (2008).

\bibitem{Coz07} M. Cozzini, P. Giorda, and P. Zanardi, Phys. Rev. B {\bf 75}, 
014439 (2007).

\bibitem {Nig90} M. P.~Nightingale, in {\em Finite-Size Scaling and Numerical 
Simulation of Statistical Systems}, edited by V. Privman, World Scientific, 
Singapore, 1990.

\bibitem{Nig93} M. P.~Nightingale, V. S.~Viswanath, and G.~M{\''u}ller, 
Phys. Rev. B {\bf 48}, 7696 (1993).

\bibitem{Bon06} O. F. de Alcantara Bonfim and J. Florencio, 
Phys. Rev. B 74, 134413 (2006).

\bibitem{Boe14} B. Boechat, J. Florencio,  A. Saguia, O. F. de Alcantara 
Bonfim, Phys. Rev. E {\bf 89}, 032143 (2014).

\bibitem{Bon14} O. F. de Alcantara Bonfim, A. Saguia, B. Boechat, 
J. Florencio, Phys. Rev. E {\bf 90}, 032101 (2014).

\bibitem{You11} W.L.~You and Y. L.~Dong, Phys. Rev. B {\bf 84}, 174426 (2011).


\bibitem {Sch09} D.~Schwandt, F.~Alet, and S.~Capponi, Phys. Rev. Lett. {\bf 103}, 170501 (2009).

\bibitem{Alb10} A.F.~Albuquerque, F.~Alet, C.~Sire, and S.~Capponi, 
Phys. Rev. B {\bf 81},064418 (2010).


\end{thebibliography}
\end{document}